\documentclass[a4paper,fleqn]{cas-sc}

\usepackage[authoryear]{natbib}
\usepackage{graphicx}
\usepackage{wrapfig}
\usepackage {ulem} 
\usepackage{subfig}
\usepackage{calc}
\usepackage{soul}
\usepackage{tikz}
\usepackage{float}
\restylefloat{table}
\usepackage{diagbox}
\usepackage{booktabs}
\usepackage{accents}
\usepackage{subfig}
\usepackage{hyperref}

\definecolor{galah1}{rgb}{0.56, 0.19, 0.19}
\definecolor{galah2}{rgb}{0.85, 0.28, 0.38}
\definecolor{galah3}{rgb}{0.94, 0.66, 0.75}
\definecolor{galah5}{rgb}{0.66, 0.66, 0.66}
\definecolor{bubble_gum}{rgb}{0.99, 0.76, 0.8}
\definecolor{timberwolf}{rgb}{0.86, 0.84, 0.82}
\definecolor{lavender_gray}{rgb}{0.77, 0.76, 0.82}
\definecolor{patriarch}{rgb}{0.5, 0.0, 0.5}	
\definecolor{darkspringgreen}{rgb}{0.09, 0.45, 0.27}
\definecolor{silver}{rgb}{0.75, 0.75, 0.75}
\definecolor{pistachio}{rgb}{0.58, 0.77, 0.45}
\definecolor{wisteria}{rgb}{0.79, 0.63, 0.86}

\def\tsc#1{\csdef{#1}{\textsc{\lowercase{#1}}\xspace}}
\tsc{WGM}
\tsc{QE}
\tsc{EP}
\tsc{PMS}
\tsc{BEC}
\tsc{DE}

\begin{document}

\title [mode = title]{Capturing episodic impacts of environmental signals} 
\shorttitle{IMPIT indices}

\author[1]{Mendiolar, M.}[orcid = 0000-0002-8380-819X] \ead{m.mendiolar@uq.edu.au}

\author[1]{ Filar, J. A.}
\author[1]{ Yang, W.-H.}
\author[2]{ Leahy, S.}
\author[3]{ Courtney, A. J.}

\address[1]{School of Mathematics and Physics, University of Queensland, St Lucia QLD 4072, Australia}
\address[2]{Queensland Department of Agriculture and Fisheries, Agri-Science Queensland, Northern Fisheries Centre, Cairns, QLD 4870, Australia}
\address[3]{Queensland Department of Agriculture and Fisheries, Level B1, Ecosciences Precinct, Joe Baker St., Dutton Park QLD 4102, Australia}

\cortext[cor1]{Corresponding author}

\begin{highlights}
    \item We developed a generic, parametrised, family of weighted indices extracted from observations of an environmental signal containing discrete episodes.
    \item The methodology considers the intensity, memory, persistence, intermittence and timing of the discrete episodes, leading to the acronym IMPIT indices.
    \item We illustrated the effectiveness and possible uses of IMPIT indices in the context of fishery and agricultural applications. 
    \item We developed IMPIT$-a$, an app that expedites the index construction and calibration processes.
    \item We expect IMPIT indices will constitute a useful new tool in the exploratory data analysis toolbox.  
\end{highlights}

\begin{abstract}
Environmental scientists frequently rely on time series of explanatory variables to explain their impact on an important response variable. However, sometimes, researchers are less interested in raw observations of an explanatory variable than in derived indices induced by episodes embedded in its time series. Often these episodes are intermittent, occur within a specific limited memory, persist for varying durations, at varying levels of intensity, and overlap important periods with respect to the response variable. We develop a generic, parametrised, family of weighted indices extracted from an environmental signal called IMPIT indices. To facilitate their construction and calibration, we developed a user friendly app in Shiny R referred to as IMPIT$-a$. We construct examples of IMPIT indices extracted from the Southern Oscillation Index and sea surface temperature signals. We illustrate their applications to two fished species in Queensland waters (i.e., snapper and saucer scallop) and wheat yield in New South Wales.
\end{abstract}


\maketitle

\section{Introduction}\label{sec_intro}\vspace{0.5em}

In many environmental studies, researchers rely on the time series of one or more relevant explanatory variables to explain variation in the response variable of interest. Frequently, the explanatory variables can be used directly, or after relatively minor modifications (e.g., transformations, lags). In such cases the essential information needed to answer the question of scientific interest is clearly reflected in the raw signal data. 

However, in some environmental applications, researchers are especially interested in indices derived from certain intermittent  episodes embedded in the time series of an explanatory variable. Most of the time they may be interested only in episodes occurring within a specified, limited, memory persisting for long enough, at sufficient intensity and occurring in sufficiently timely manner with respect to the response variable. In our context, this limited memory will be a period of fixed and uninterrupted duration in the past, with respect to the current observation time.

Consider, for example, the time series of the Southern Oscillation Index (SOI), a signal that is widely utilised in environmental sciences. SOI indicates the magnitude and direction of the El Ni\~no southern oscillation (ENSO), in which El Ni\~no (sufficiently negative and sustained SOI values) typically results in hotter and drier than average conditions in Australia, while La Ni\~na (sufficiently positive and sustained SOI values) is associated with cooler and wetter than average conditions in Australia. Similar considerations would apply in the case of marine heatwaves (MHWs) \citep{Barbeaux_2020marine, Yao_2021variations}, and associated episodes that would be extracted from the time series of Sea Surface Temperature (SST). However, high/low SOI value episodes and marine heatwaves occur intermittently, persist for varying duration, and differ in intensity. Furthermore, if the research focuses on their influence on the harvest yield, we may well wish to identify those episodes that overlap key life history stages of that species or the crop. For fisheries, the spawning season is clearly a sensitive life stage. In the agricultural context, we could consider the pre-sowing, sowing or flowering seasons. 

Moreover, if the expert domain knowledge indicates that there is only a limited time horizon over which the analysis is meaningful. For simplicity, we call that horizon a {\it memory}. 
This immediately leads to the problem of appropriately differentiating between episodes that occur more or less recently, within that memory, which we refer to as {\it recency}.

In this paper, we develop a generic, parameterised family of weighted indices extracted from observations of an environmental signal, on the basis of intermittent episodes of interest embedded in the signal. These weighted indices can be calibrated to capture the relative influence of intermittency, memory, persistence, intensity and timing of the underlying episodes. Hence, we name them {\it IMPIT indices}. To facilitate ease of construction and calibration of IMPIT indices, we developed a user friendly open source app called IMPIT$-a$. 

We illustrate the effectiveness of the design and calibration of IMPIT indices with fishery catch rate data from two fished species in Queensland, snapper (\textit{Chrysophrys auratus}) and saucer scallop (\textit{Ylistrum balloti}), and agricultural yield data of wheat production in New South Wales. We consider correlations between certain environmental variables and standardised catch per unit effort (SCPUE) and yield per hectare as response variables. We show that highly significant correlations arise between these and IMPIT indices extracted from SOI and/or SST signals, despite analyses indicating no significant (or only weak) correlation between these response variables and the raw signal data.

To the best of our knowledge, prior studies of discrete, intermittent episodes in the environmental and ecosystem context are very application specific. See, for instance, \citet{Brad_2003proxy, Ensminger_2004intermittent, Unal_2013summer, Bakun_2014active, Strydom_2020too, Bellanthudawa_2022spectral}. By contrast, we construct a generic parametrised family of indices (i.e., IMPIT) induced by episodes embedded in the time series of an environmental signal. Moreover, we propose a systematic parametric search for associations between these indices and seemingly exogenous response variables, such as SCPUE of harvested fish species and yield per hectare of crop production. Importantly, the proposed IMPIT indices can be fine-tuned to capture a wide range of cumulative effects of these discrete episodes.

Since discrete intermittent episodes are embedded in the time series of an environmental signal - for the sake of completeness - we next briefly differentiate our approach from a large body of existing time series and signal processing techniques. The majority of time series methods focus on forecasting episodes of interest. For instance, the well-known Croston method (see \citet{Croston_1972forecasting}) was designed for forecasting intermittent demands. However, it is known that for many intermittent episodes (e.g., earthquakes, or El Ni\~no events) such forecasting can be very challenging \citep{Rundle_2021complex, Ham_2019deep, Glantz_2015shades, Ludescher_2014very}. 

Sophisticated signal processing techniques are also used to extract features of signals and build indices accordingly. Popular among these are techniques exploiting Empirical Orthogonal Functions (EOFs), Fourier transforms, and wavelet transforms \citep{Jolliffe_2011PCA, Hannachi_2007empirical, Thomson_2014data}. Specifically, several climate indices are built on EOFs such as the Arctic Oscillation \citep{Thompson_1998arctic} and RMM1 and RMM2 indices for monitoring the Madden-Julian Oscillation \citep{Wheeler_2004all}. Moreover, the features extracted by these methods are used as explanatory variables in regression analyses to answer scientific inquiries. For example, \cite{Yang_2013ecological} utilised the EOFs and short-time Fourier transform together to reveal the relationship between spawning success of shovelnose sturgeon in the Lower Missouri River and river depth and water temperature using fish tracking data. 

An important distinction between those approaches and this study is that we are not attempting to forecast the underlying intermittent episodes. Instead, we construct a unified parametrised family of indices induced by episodes embedded in an environmental signal. We are not using signal processing methods to automatically identify the discrete intermittent influential episodes that we are working with. These are either assumed to be already well defined (such as, marine heatwaves) or are natural (such as, threshold crossing episodes). Note that IMPIT indices in this study are constructed in the natural time domain rather than the frequency domain.

Moreover, we propose a systematic parametric search for associations between the IMPIT indices and important seemingly exogenous response variables, such as SCPUE of harvested fish species and yield per hectare of crop production. Thus, the proposed IMPIT indices are driven by features of these discrete but intermittent episodes occurring within the observed signals (such as SOI or SST). They also account for the propagation effects of the intermittent episodes. The proposed search process prepares them to be plausible indicators or explanatory variables to identify and unravel their possible impacts on relevant response variables.

We also point out that the IMPIT indices are quite generic and hence have potential application in a wide class. The indices and their calibration should be seen as another tool of exploratory data analyses. Indeed, they have recently been used in a real-world application \citep{Filar_2021modelling}. Finally, introduction of IMPIT indices is also timely because of recent and ongoing changes in the intensity and frequency of extreme weather and climate events \citep{Ummenhofer_2017extreme, Frolicher_2018marine, Oliver_2019mean}. 

The paper is structured as follows. Section 2 describes the methodology used to develop IMPIT indices. Section 3 illustrates their design and calibration for two widely used environmental signals and study their impact in a fishery and agricultural context. Section 4 describes the app developed and its visualization capabilities. Finally, Section 5 contains a brief summary and discussion.

\section{Indices capturing Intermittence, Memory, Persistence, Intensity and Timing (IMPIT) \label{sec_IMPIT}}\vspace{0.5em}

We consider a time series of an environmental signal denoted by $\mathbf{X}=\{X_t\}$ with time $t=T, \dots, 1$, where $X_T$ represents the earliest observation and $X_1$ the latest (most recent) observation. The units of time should be consistent among signal points and can be hours, days, years, etc. We define the {\it memory} as a period of fixed and uninterrupted duration and denote by $m$ its length. With the memory fixed, we can extract a sequence of environmental episodes where each episode satisfies certain characterising conditions. 
The impact of these episodes on a response variable may depend on their intermittence, persistence, intensity and timing. Timing refers to the position of the episode in the time history with respect to any special aspects of the studied phenomenon (e.g., spawning or flowering seasons). 
Whatever their definition, the essential requirement of the episodes is that there is some number $K \leq m$ of them and, that the entire remembered history of the signal $\mathbf{X}=\{X_1,X_2, \ldots, X_m \}$ can be sub-partitioned into these $K$ mutually exclusive episodes $E_k=\{ X_{s_k}, X_{s_{k+1}}, \ldots, X_{s_{k+n_k-1}} \}$, $k=1, \ldots, K$, where $n_k$ denotes the length of the $k^{th}$ episode and $s_k$ is its starting time location. Namely,
\begin{equation}\label{Eq_sub_part}
    \mathop{\cup}_{k = 1}^{K} \left[ E_k \right] \subseteq \mathbf{X}. 
\end{equation}
Note that while the elements comprising the episode $E_k$ would most often be consecutive, this is not a requirement. Thus, it is possible that $X_{s_{k+1}} \ne X_{s_{k}+1}$. However, in most of our illustrations the episodes indeed consist of consecutive observations and in such a case, they will be denoted by $E_k=\{ X_{s_k}, X_{s_k+1}, \ldots, X_{s_k+n_k-1} \}$, $k=1, \ldots, K$. 

The intensity, $I(E_k)$, of the episode $E_k$ can be any function of the observations comprising $E_k$, as explained in the next Subsection \ref{subsec_intensity}. Collectively, the set of intensity values will be denoted by the symbol $\boldsymbol{I}$. Furthermore, we need to capture the relative importance of the episodes $E_1, \ldots, E_K$. This will be done with the help of relative importance weights, collectively denoted by $\boldsymbol{w}$. Note that the intensity is a measure of the strength of each episode, whereas importance weights are intended to capture the relative impact of multiple episodes on the phenomenon of interest.

With respect to the sub-partition in expression \eqref{Eq_sub_part}, we define a family of associated IMPIT indices extracted from the signal of the environmental variable $\mathbf{X}$. This family is of the form
\begin{equation}\label{Eq_X_ind}
    X(\boldsymbol{I},\boldsymbol{w}) = \sum_{k=1}^{K} w(E_k)\textit{I}(E_k),
\end{equation}
where $w(E_k)$ and $I(E_k)$ denote the importance weight and the intensity measures associated with episode $E_k$, respectively.

\subsection{Intensity of episodes\label{subsec_intensity}}\vspace{0.5em}
An intensity function $I(E_k)=I(X_{k_1}, X_{k_2}, \ldots, X_{k_{n_k}})$ should map observed values in the episode onto a number capturing the intensity or strength of that episode. Several natural candidates could be considered for such functions including, the mean of the observations $\{X_{k_1},X_{k_2}, \ldots, X_{k_{n_k}} \}$
\begin{equation}\label{Eq_I_av}
    I(E_k) = \bar{I}(E_k) = \frac{1}{n_k} \sum_{j=1}^{n_k} X_{k_{j}},
\end{equation}
or the logarithm of their sum 
\begin{equation}\label{Eq_I_ln}
    I(E_k) = I^{l}(E_k) = log \left( \sum_{j=1}^{n_k} X_{k_{j}} \right).
\end{equation}
Other candidates such as, the median, geometric mean, signal-to-noise ratio could also be considered (ensuring that each is well defined). 

\subsection{Weights of episodes \label{subsec_weights}}\vspace{0.5em}

An essential ingredient in the construction of an IMPIT index (i.e., Equation \eqref{Eq_X_ind}) is the relative importance weight of each episode (i.e., $w(E_k)$ for each $k$). We postulate that the weights $w(E_k)$ are all non negative numbers lying in the interval $[0,1],$ with $1$ corresponding to episodes viewed as most important to the studied phenomenon in the aggregate index $X(\boldsymbol{I},\boldsymbol{w})$.
There are, at least, the following three approaches to the construction of these weights: 
\begin{enumerate}
    \item[(1)] Multiplicative, product form, of these weights given by%
    \begin{equation}\label{Eq_w}
        w(E_k) = w_1(n_k,m) w_2(s_k,m) w_3(E_k),
    \end{equation}
    where $w_1, w_2, w_3 \in [0,1]$ are intended to capture the importance of \textit{persistence}, \textit{recency} and \textit{timing} of the episode $E_k$ of length $n_k$ and starting position $s_k$ in the memory of length $m$. Any of these weights could also be set to $0$ or $1$ depending on exogenous information available. In absence of deeper understanding of the persistence, recency and/or memory aspects, $w_1(n_k,m)$, $w_2(s_k,m)$ and $w_3(E_k)$ can be set equal to $1$ for every $k$.
    \item[(2)] Construction where the weights can be used purely as a technical tool to achieve a desired form of the $X(\boldsymbol{I},\boldsymbol{w})$ indices. This is illustrated in the remark, below.
    \item[(3)] Construction based on expert domain knowledge. This may depend on either more detailed understanding of the phenomena defining the episodes, or on the intended target response variable that a study aims to explain (at least partially) with the help of the $X(\boldsymbol{I},\boldsymbol{w})$ indices\footnote{We do not discuss this approach in any more detail as our applications are used mainly as illustrations of the generic methodology.}.
\end{enumerate}    

{\bf Remark:} Note that most standard statistical indices can be easily recovered within the above $X(\boldsymbol{I},\boldsymbol{w})$ family. For instance, if we wanted the mean of observations $\{X_1, X_2, \ldots, X_{12} \}$, all we need to do is define $E_k = X_k$ for each $k$, the memory $m=12$, the intensity function to be $I(E_k)=\frac{X_k}{12}$ and all the weights to be $w(E_k)=1$, for each $k$. Similarly, other indices such as medians, moving averages, or coefficient of variation can be naturally constructed in the above form.

In the remainder of this subsection, we propose certain specific algebraic forms for the weights in \eqref{Eq_w} which require users to calibrate only a small number of parameters  while offering considerable freedom in choosing their shape.

\subsubsection{Persistence \label{subsubsec_persistance}}\vspace{0.5em}

\begin{figure}[ht]
    \centering
    \includegraphics[width=6cm]{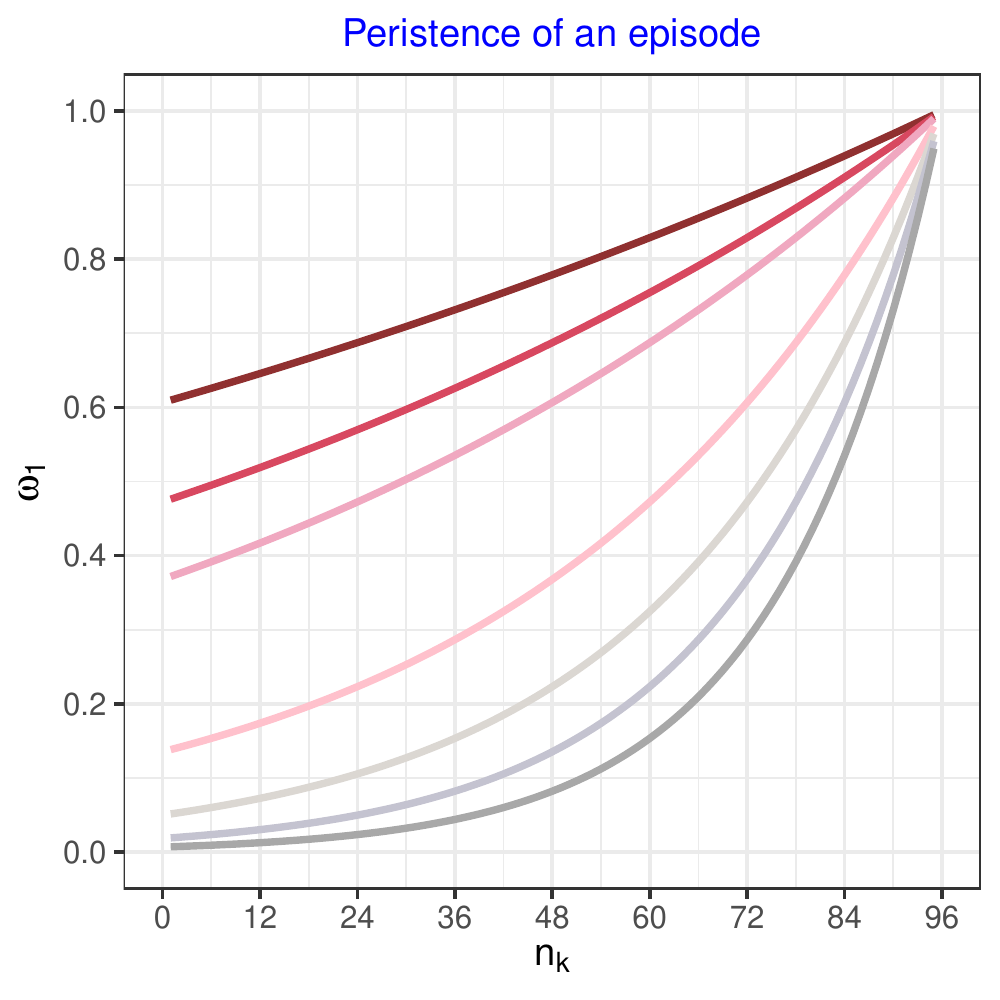}
    \caption{Examples of the shape of $w_1(n_k,m)$ for different dampening parameters $a$, where $n_k$ is measured in months. Colour code: $a=0.50$ (\textcolor{galah1}{\rule[0.23em]{1em}{0.15em}}), $0.75$ (\textcolor{galah2}{\rule[0.23em]{1em}{0.15em}}), $1.00$ (\textcolor{galah3}{\rule[0.23em]{1em}{0.15em}}), $2.00$ (\textcolor{bubble_gum}{\rule[0.23em]{1em}{0.15em}}), $3.00$ (\textcolor{timberwolf}{\rule[0.23em]{1em}{0.15em}}), $4.00$ (\textcolor{lavender_gray}{\rule[0.23em]{1em}{0.15em}}) and $5.00$ (\textcolor{galah5}{\rule[0.23em]{1em}{0.15em}}).}
    \label{fig:w1}
    \vspace{-2em}
\end{figure}

It is well known that persistence of certain environmental events can have considerable impact. For instance \citet{Li_2021widespread} state {\it ``Persistence has a key role in the climate impacts of a given temperature event''}. To describe the persistence of an episode we used the following functional form
\begin{equation}\label{Eq_w1}
    w_1(n_k,m) = exp \left( -a \left( 1 - \frac{n_k}{m} \right) \right), 
\end{equation}
where $n_k$ is the length of episode $E_k$, $m$ the memory, and $a>0$ the dampening parameter. For each fixed $a$, this formula places low weights on short episodes (see Figure~\ref{fig:w1}). This figure also shows that different values of parameter $a$ control both the slope and the curvature of the parameterised family of convex functions represented by \eqref{Eq_w1}.

\subsubsection{Recency \label{subsubsec_recency}}\vspace{0.5em}

Recency of an episode could also play an important role in its impact on the studied phenomena \citep{Deryugina_2013people, Hoffmann_2022climate}. Frequently, but not always, recent episodes can be expected to have greater impact than those that happened in more distant past. An obvious indicator of the recency of the episode $E_k=\{ X_{s_k}, X_{s_k+1}, \ldots, X_{s_k+n_k-1} \}$ is the ratio of $s_k/m \in [0,1]$, with high values indicating less recent events. If we use this ratio to design an importance weight that calibrates recency, we could use the following two-step process.
 
\begin{figure}[ht]
    \centering
    \includegraphics[width=10.5cm]{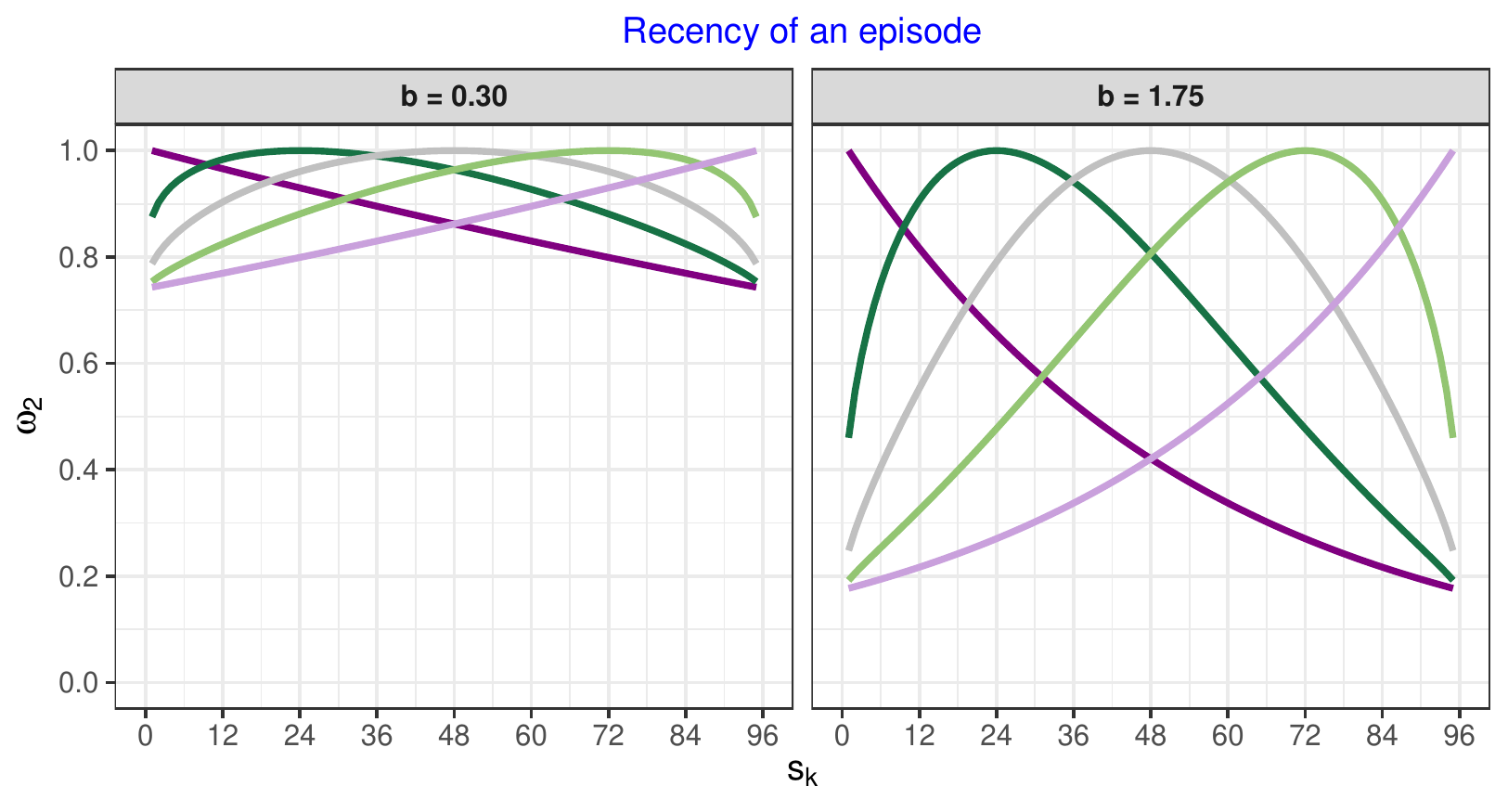}
    \caption{Examples of the shape of $w_2(s_k,m)$ function when $m=96$ (months) and parameters $c$ and $b$ take the indicated values. Colour code: $c=0.00$ (\textcolor{patriarch}{\rule[0.23em]{1em}{0.15em}}), $0.25$ (\textcolor{darkspringgreen}{\rule[0.23em]{1em}{0.15em}}), $0.50$ (\textcolor{silver}{\rule[0.23em]{1em}{0.15em}}), $0.75$
    (\textcolor{pistachio}{\rule[0.23em]{1em}{0.15em}}) and $1.00$ (\textcolor{wisteria}{\rule[0.23em]{1em}{0.15em}}).}
    \label{fig:w2}
    \vspace{-1em}
\end{figure}

Let us define the function of $s_k$ (starting time of the episode $E_k$) and $m$ the memory by 
\begin{equation}\label{Eq_nu}
    \nu(s_k,m) = \lambda \left( \frac{s_k}{m} \right)^{c} \left( 1- \frac{s_k}{m} \right)^{1-c},
\end{equation}
where $0 \leq c \leq 1$, and the scaling parameter $\lambda > 0$ is chosen to ensure that $\nu(s_k,m) \leq 1$. To capture the relative importance of recency of episode $E_k$ within the period considered we again propose an exponential form
\begin{equation}\label{Eq_w2}
    w_2(s_k,m) = exp \left[ -b \left( 1- \nu(s_k,m)\right) \right],
\end{equation}
where $b>0$. When Equation \eqref{Eq_nu} is substituted into \eqref{Eq_w2}, this ensures that $w_2(s_k,m)$ also range between $0$ and $1$ and attain a maximum when $\nu(s_k,m)=1$. The parameter $b$ can be viewed as a dampening (or accelerating) factor of the rate of decay away from the maximum. For instance, from Figure~\ref{fig:w2} we see that $b=0.3$ dampens the rate of decay of these weights to always remain above $0.7$, while $b=1.75$ permits most of them to drop to below $0.3$.

The parameter $c$ can be used to assign more or less importance to episodes that occur earlier or later in the history. From Figure~\ref{fig:w2} we can see that when $c=0$ (dark purple line), episodes starting late in the memory ($s_k$ close to $m$) obtain $w_2(s_k,m)$ values that are smaller than episodes starting early in that history ($s_k$ close to 1), and conversely when $c=1$ (light purple line). Note that the peak of $w_2(s_k,m)$ coincides with the value of $s_k$ that maximises $\nu(s_k,m)$, and hence also $w_2(s_k,m)$. For any fixed $c$ strictly between $0$ and $1$, $s^*_k = c m$ is the maximiser of $\nu(s_k,m)$. Since the starting time is an integer, in practice, the peak will be set to occur at the nearest integer to that ratio. When parameters $a$, $b$ and $c$ are calibrated to optimise a performance measure, researchers may be particularly interested in the ratio $c = \frac{s^*_k}{m}$ as it is an indicator of the most important relative recency, with respect to the memory $m$.

\subsubsection{Timing \label{subsubsec_timing}}\vspace{0.5em}

Depending on the application, the importance weights of episodes may need to be modulated further by the timing of the occurrence of these episodes with respect to one or more response variables. For instance, in the fishery science application, discussed below, the response variable is the SCPUE, which is a proxy for abundance of the species of interest. If the environmental episodes of the IMPIT indices (e.g., MHWs) occurred at times overlapping sensitive periods in the species' lifecycle (e.g., spawning season), that information should modulate the values of the indices. 
Mathematically, such modulation can be modelled in a variety of ways. Perhaps, the easiest of these is by incorporating a third, multiplicative, importance weight $w_3 \in [0,1],$ designed to capture the special timing of interest (if any) of the occurrence of the episode $E_k.$

We shall denote the special timing we are interested in by $\mathscr{T}$. Hence, the overlap with the episode $E_k$ can be denoted by $E_k \cap \mathscr{T} = \mathscr{T}_{k}$ and its length by $\tau_k$. Now we define the third {\it timing} weight by 
\begin{equation}\label{Eq_w3}
    w_3(E_k) = 1-exp\left[-d\left(\frac{\tau_k}{n_k} \right)\right],
\end{equation}
where the fraction $\tau_k / n_k$ is intended to capture how much of the episode is taken by this overlap and $d$ is a dampening parameter, $d \geq 0$. 

\begin{figure}[ht]
    \centering
    \includegraphics[width=15.5cm]{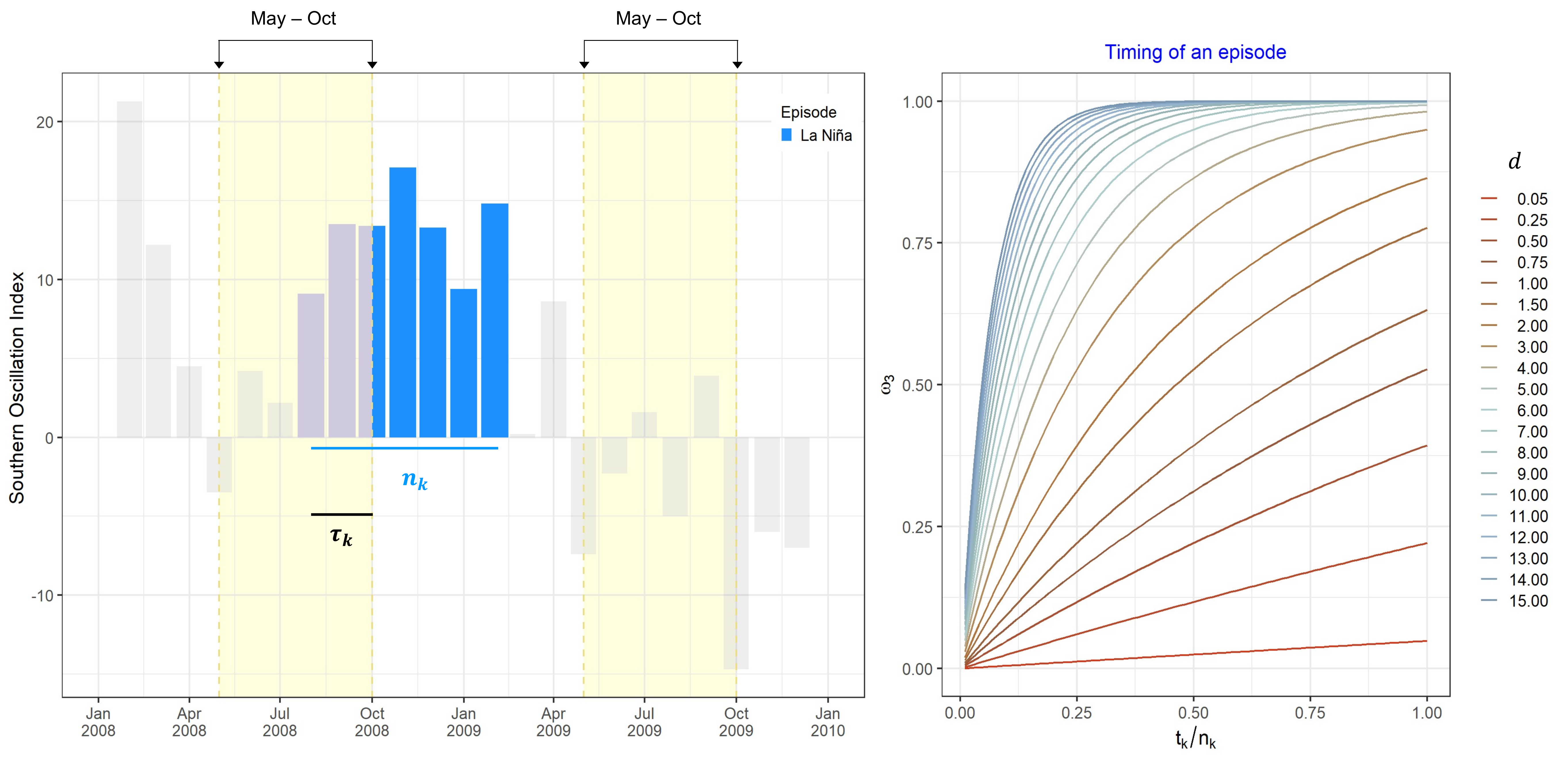}
    \caption{Left panel: Example of overlapping between La Ni\~{n}a episode during 01/2008-01/2010 and the May-October spawning season for snapper (\textit{Chrysophrys auratus}) in yellow. Y-axis indicates monthly SOI value, with sustained SOI values above $8$ for at least $5$ consecutive months indicating a La Ni\~na episode. Right panel: Examples of the shape of $w_3(E_k)$ function when $m=96$. The differences in color correspond to changes in the parameter $d$.}
    \label{fig:w3}
    \vspace{-1em}
\end{figure}

The left panel of Figure~\ref{fig:w3} illustrates a typical overlap of the special timing of the spawning season of snapper with a $2008$ La Ni\~na episode. Snapper spawn in aggregations over several months (generally May to October) and synchronise spawning on the lunar cycle \citep{Wortmann_2018snapper}. The right panel of Figure~\ref{fig:w3} shows the impact of the dampening parameter $d$ on the shape of $w_3(E_k),$ as a function of the ratio $\tau_k / n_k$. 

\subsection{Periodic seasonality indices \label{subsec_period}}\vspace{0.5em}

In this case each episode $E_k$ is of equal length $n_k=n$ and they occur with periodic regularity. 
\begin{equation}
    E_k = \{X_{k_1}, X_{k_2}, \ldots, X_{k_n} \}, \; \; \; k = 1, \ldots, K.
\end{equation}
where the number $K$ of these episodes is limited by spacings between them and the total memory $m$ under consideration. For instance, consider a memory of, say, $m=48$ months counting backwards from December of the current year. For episodes capturing the spawning aggregations of snapper (May-October) there would be exactly $K=4$ such episodes, consisting of $ E_1=\{X_3,X_4, \ldots, X_8 \}, E_2=\{X_{15}, X_{16}, \ldots, X_{20} \}, E_3=\{X_{27}, X_{28}, \ldots, X_{32} \}$ and $E_4=\{X_{39}, X_{40}, \ldots, X_{44} \}$. 

\subsection{Threshold-crossing indices \label{subsec_threshold_cross}}\vspace{0.5em}

These concern situations where episodes of interest are defined by only those observations which exceed, or fall below some specified threshold for sufficiently long.
Let $\delta$ be a threshold and $E_k$ be the $k^{th}$ episode of interest, of duration $n_k$, be defined by either
\begin{equation}\label{Eq_up_epi}
    E_k = E^{u}_k (\delta, \ell) = \{ X_{k_j} | X_{k_j} \ge \delta, j=1,\ldots, n_k \; \; \& \; \; n_k \ge \ell \},
\end{equation}
or
\begin{equation}\label{Eq_dwn_epi}
    E_k = E^{d}_k (\delta,\ell) = \{ X_{k_j} | X_{k_j} \le \delta, j=1,\ldots, n_k \; \; \& \; \; n_k \ge \ell \},
\end{equation}
where $\ell$ denotes the minimum required duration (with $\ell = 1$ being the default). Here, the superscript $u$ denotes up-episodes (above the threshold) whereas $d$ denotes down-episodes (below the threshold). The selection of the sub-sequences of the $n_k$ observations $X_{k_j}$ comprising $E_k$ would typically depend on additional, application specific knowledge. 

Next, any IMPIT index of the form \eqref{Eq_X_ind} associated with episodes $E^{u}_k (\delta, \ell)$ will be called a Super$X^{u} (\delta, \ell)$ index. Similarly, any IMPIT index associated with episodes $E^{d}_k (\delta, \ell)$ will be called a Sub$X^{d} (\delta, \ell)$ index. In the default case of $\ell=1$, we simplify the above notation to $E^{u}_k (\delta)$ and Super$X^{u} (\delta)$ (respectively, $E^{d}_k (\delta)$ and Sub$X^{d} (\delta)$). Since increasing $\ell$ only makes conditions in \eqref{Eq_up_epi} more restrictive, it follows that the number of 
$E^{u}_k (\delta)$ episodes is greater or equal than that of $E^{u}_k (\delta, \ell)$ episodes, for any $\ell >1$. Similarly for \eqref{Eq_dwn_epi} and the down episodes.

\begin{figure}[ht]
    \centering
    \includegraphics[width=15cm]{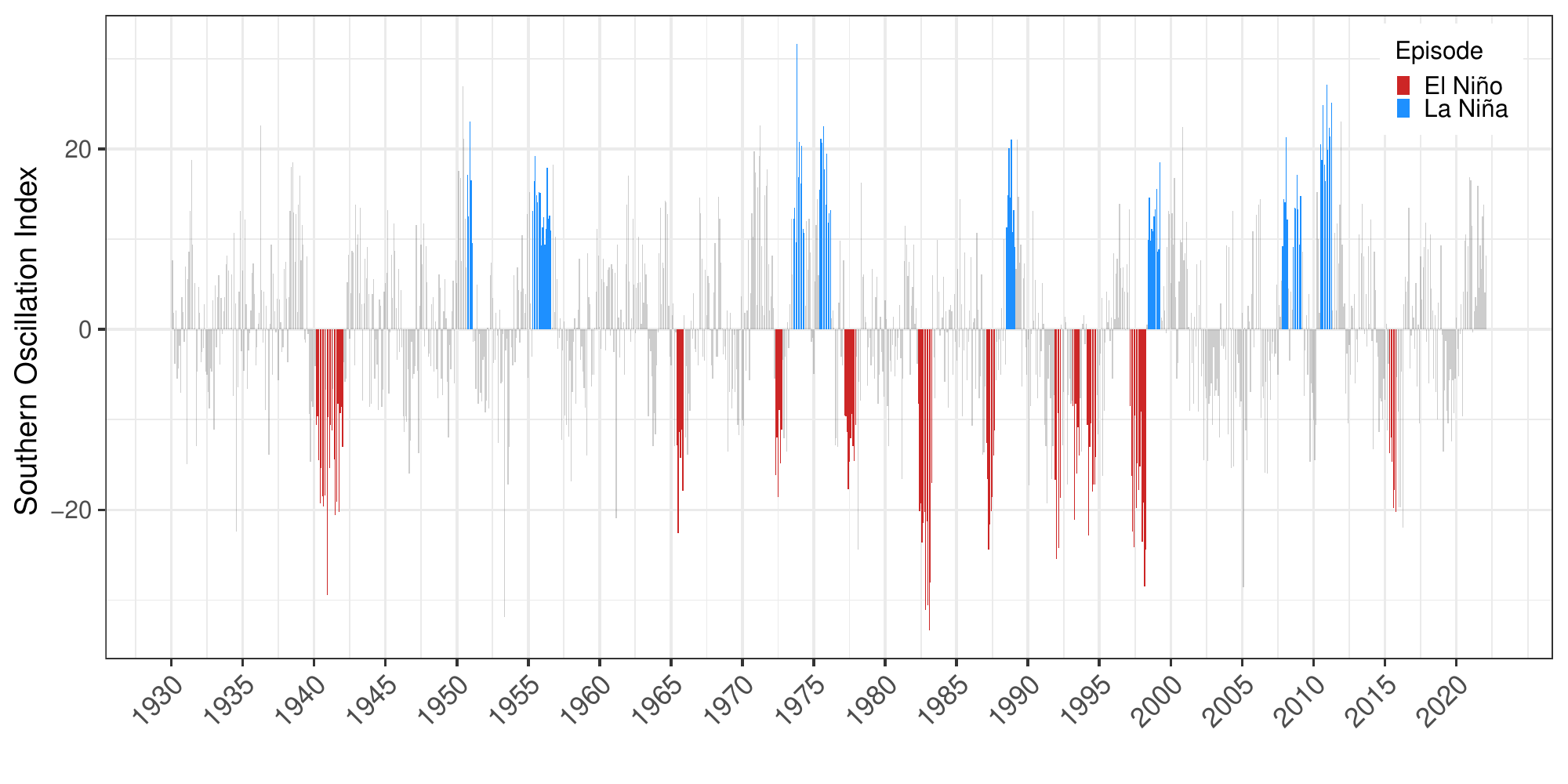}
    \caption{Time series of the monthly Southern Oscillation Index (SOI) from January 1930 to December 2020. Based on \citet{Bureau_2012record} definition for El Ni\~no and La Ni\~na episodes, values below -8 and above 8 for at least five consecutive months are shaded red and blue, respectively.}
    \label{fig:soi_1930-2020}
    \vspace{-1em}
\end{figure}

The extensively studied El Ni\~no and La Ni\~na episodes illustrate this situation. They are extracted from the SOI time series $\mathbf{X}$. In particular, according to \citet{Bureau_2012record}, an El Ni\~no episode is simply $E^{d}_k(-8,5)$ where the threshold $\delta=-8$ and minimum duration of $\ell=5$ are used. Analogously, a La Ni\~na episode corresponds to $E^{u}_k(8,5)$ (see Figure~\ref{fig:soi_1930-2020}). 

\subsection{Calibration via exploration of the parameter space \label{subsec_calibration}}\vspace{0.5em}

An underlying contribution of this study is the demonstration that importance weights of the episodes $E_k$ may be calibrated to achieve specific research objectives. In our application, the dual objective is to identify parameter configurations that: (a) achieve high absolute value of the Pearson correlation coefficient between an IMPIT index and a response variable meaningful in fish stock assessment, and (b) are sufficiently stable with respect to small changes in other parameters, especially the memory. 

For each, fixed, memory parameter $m$, all the weights discussed above are fully characterised by the choice of the parameter vector $(a,b,c,d),$ where $a,b,d \ge 0$ and $c \in [0,1].$ In most applications, the parameters $a$, $b$ and $d$ will also have practical upper bounds. Thus, numerically, the search for an optimal parameter configuration can be carried out by a grid search of a $4-$dimensional hyperrectangle. That search could then be repeated for multiple values of the memory parameter $m$. With respect to objective (a), an approximately optimal configuration $(m^*, a^*,b^*,c^*,d^*)$ could be identified by such an exhaustive search. 

However, such a brute-force calibration has some drawbacks. Clearly, it is computationally intensive. Further, the best parameter configuration with respect to (a) may be unstable with respect to (b) and may result in counter-intuitive importance weights, in the context of the application.

Hence, in our illustrative application, we propose the following {\it stage-wise explorative calibration} that is simple to implement, preserves the logical order of the design of IMPIT indices, and is likely to identify multiple interesting parameter configurations for further consideration.

At the first stage, for each value of memory $m$, plots of correlation coefficient on the vertical axis ($y$-axis) and $a$ values on the horizontal axis ($x$-axis) are generated while setting $w_2=w_3=1$. Based on these plots a candidate value of the parameter $a$ is chosen. This choice determines the $w_1$ weight and may, but need not necessarily, correspond to the maximum of the absolute value of the relevant correlation \footnote{Stability of the correlation coefficient with respect to changes in $a$ is also taken into consideration. If the maximization criterion does not meaningfully differentiate among $a-$values, expert suggested value (in the range) can be selected.}. At the second stage, with the already chosen $a-$value and $w_3=1$ fixed, plots of correlation coefficient on the vertical axis ($y$-axis) and $c$ values on the horizontal axis ($x$-axis) are generated, for alternative pairs of $m$ and $b$ parameters. These plots can be visualized in a composite ``map", in the parameter space, in order to explore and find candidate parameter configurations that meet both objectives (a) and (b). At the third stage, we repeat the preceding while considering only episodes that overlap a special timing season. This stage consists of generating composite maps for alternative values of $d$ and selecting promising candidate configurations.

We note that even in the absence of a response variable suggested by a specific application, IMPIT indices can still be optimally calibrated with respect to their own desirable characteristics. For instance, for each fixed $m$, configuration $(a,b,c,d)$ can be chosen to maximize the $R^2$ statistic of the corresponding IMPIT index signal when fitted to a prescribed model (e.g., a linear model, or a power law). Moreover, the strength of the influence can be captured by a wide range of quantitative measures (e.g., goodness-of-fit or efficiency measures), see for instance \citet{Krause_2005comparison}. The choice of such a measure, while important, is not the focus of this study. 

\section{Case studies using fishery and agricultural data
\label{sec_applications}}\vspace{0.5em}

We illustrate the benefits of the IMPIT indices' design and calibration in the context of the impact of selected environmental signals on two fished species in Queensland, snapper (\textit{Chrysophrys auratus}) and saucer scallop (\textit{Ylistrum balloti}), and yield data of wheat production from NSW. The environmental signals are the Southern Oscillation Index (SOI) and sea surface temperature (SST). The response variables are the time series of SCPUE and yield per hectare of wheat. These signals were selected because studies have demonstrated their strong correlations with fish catch rates \citep{Joll_1995environmental, Lenanton_2009ongoing, Caputi_2019factors} and wheat production \citep{Gutierrez_2017impacts, Wan_2022does}, particularly in Australia \citep{Rimmington_1993forecasting, Yuan_2015impacts, Zheng_2018value, Potgieter_2002spatial}. A description of the data acquisition and episodes considered in these case studies is given in Section \ref{subsec_data} below.

\subsection{Data acquisition \label{subsec_data}}\vspace{0.5em}

\subsubsection*{Environmental data and two types of intermittent episodes\label{subsec_data_env}}\vspace{0.5em}

Monthly SOI values were obtained from the Australian Bureau of Meteorology (BoM). The SOI is computed from the variations of monthly mean sea level pressure difference between Tahiti and Darwin \citep{Chowdhury_2010australian}. Positive SOI values are generally associated with a La Ni\~na pattern in the central and eastern equatorial Pacific and above-average winter/spring rainfall for Australia, particularly across the east and north. Negative SOI values are associated with El Ni\~no conditions and lower than average winter/spring rainfall over much of eastern Australia \citep{Bureau_2012record}.

\begin{figure}[ht]
    \centering
    \includegraphics[width=8cm]{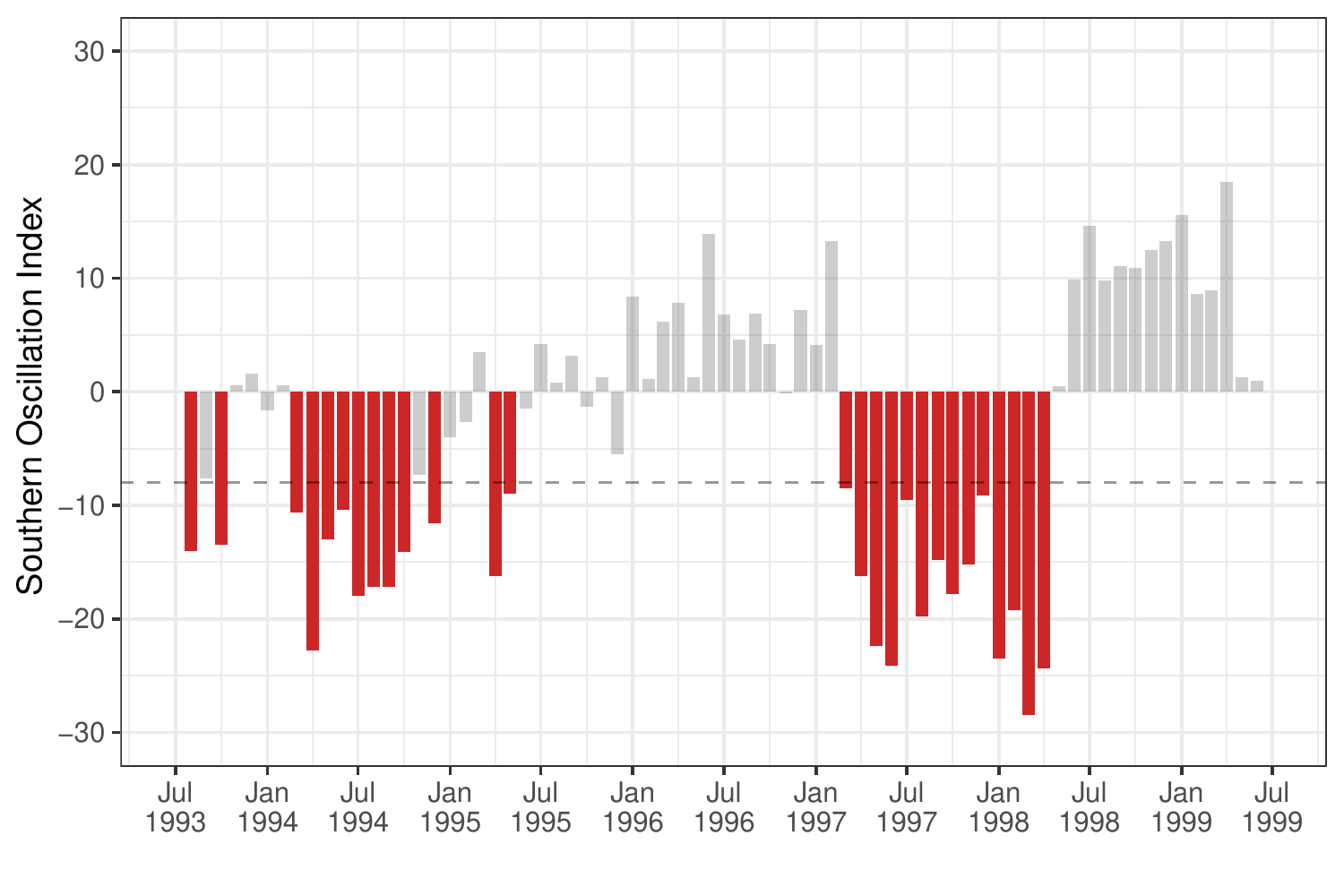}
    \includegraphics[width=8cm]{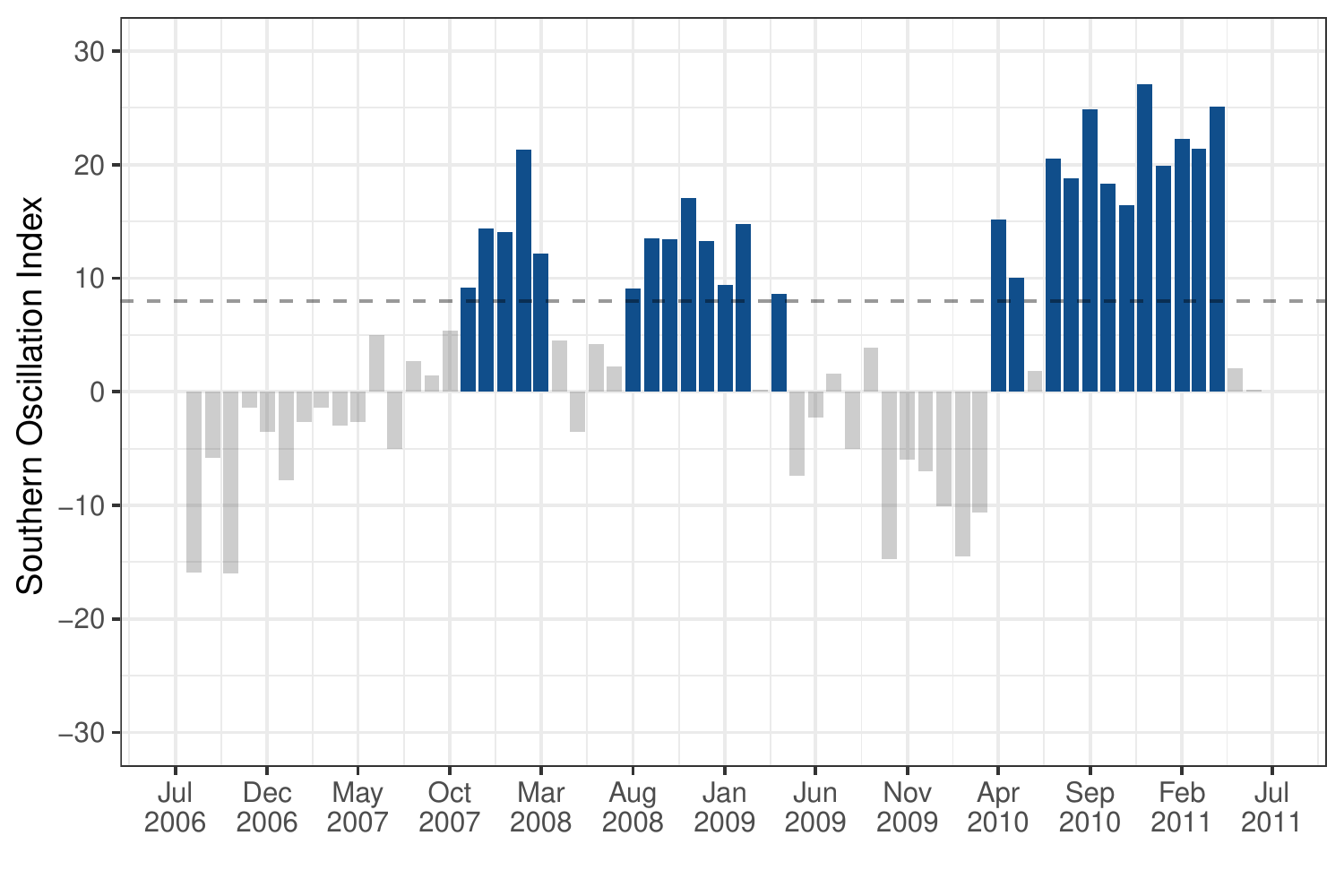}
    \caption{Monthly SOI values during $07/1993-07/1999$ and $07/2006-07/2011$. Columns in red and blue correspond to values below and above $-8$ and $8$ thresholds (dashed line) used to illustrate some of the $E^d(-8)$ and $E^u(8)$ episodes used in this study.}
    \label{fig:Eu8_episodes}
    \vspace{-1em}
\end{figure}

Let $\mathbf{X}$ denote the underlying SOI signal. We illustrate a class of threshold crossing IMPIT indices extracted from $\mathbf{X}$. In particular, we consider up-episodes $E^{u}_k (8)$ and down-episodes $E^{d}_k (-8)$ corresponding to Super$X^{u} (8)$ and Sub$X^{d} (-8)$ indices, requiring calibration. The choice of the thresholds $\delta=8$ and $\delta=-8$ was guided by their use in a common definition of La Ni\~na and El Ni\~no episodes, respectively \citep{Bureau_2012record}. 
For example, Figure~\ref{fig:Eu8_episodes} displays five $E^{u}_k (8)$ episodes during $07/2006-07/2011$ (right panel) and six $E^{u}_k (8)$ episodes during $07/1993-07/1999$ (left panel). Note that in the case of the up-episodes $E^{u}_k (8)$ only three of these correspond to La Ni\~na episodes lasting at least five months, since the single month episode in April $2009$ and then the two monthly episode in April-May $2010$ do not qualify as La Ni\~na episodes. Similarly, in the case of the down-episodes $E^{d}_k (-8)$ only two of the six correspond to El Ni\~no episodes. 

The SST data was sourced from the Integrated Marine Observing System (IMOS) database with $6$ day average, night time capture (``ghrsst\_L3S\_6d\_ngt'') over a period of approximately 28 years (01/04/1992 to 31/12/2019). The 6 day average night time data set was used to avoid any daytime temperature artefacts (e.g. sun glint). Also, 6 day averaging fills gaps that are otherwise present in the daily time series as a result of cloud cover blocking satellite view of the ocean surface. The spatial distribution of SST used in the analysis corresponded to the areas of the scallop fishery described in \cite{ONeill_2020stock}.

Marine heatwaves (MHWs) are anomalous ocean temperature events, identified by extremely warm SST that persists for days to months \citep{Hobday_2016hierarchical}. They can be caused by a mix of atmospheric forcing and oceanographic conditions and depend on location and season. For instance, a heatwave that affects coral reefs in warmer waters will have higher temperatures than one that affects kelp forests in cooler waters.

We followed the hierarchical definition of MHW episodes according to \cite{Hobday_2016hierarchical}. This definition, takes into account the following key features: (i) Anomalously warm temperatures with respect to a baseline average temperature over a period of 30 years, and a high percentile threshold of 90; (ii) Prolonged events persisting for at least five days; (iii) Discrete events appearing with sufficient separation between successive events. 

\begin{figure}[ht]
    \centering
    \includegraphics[width=8cm]{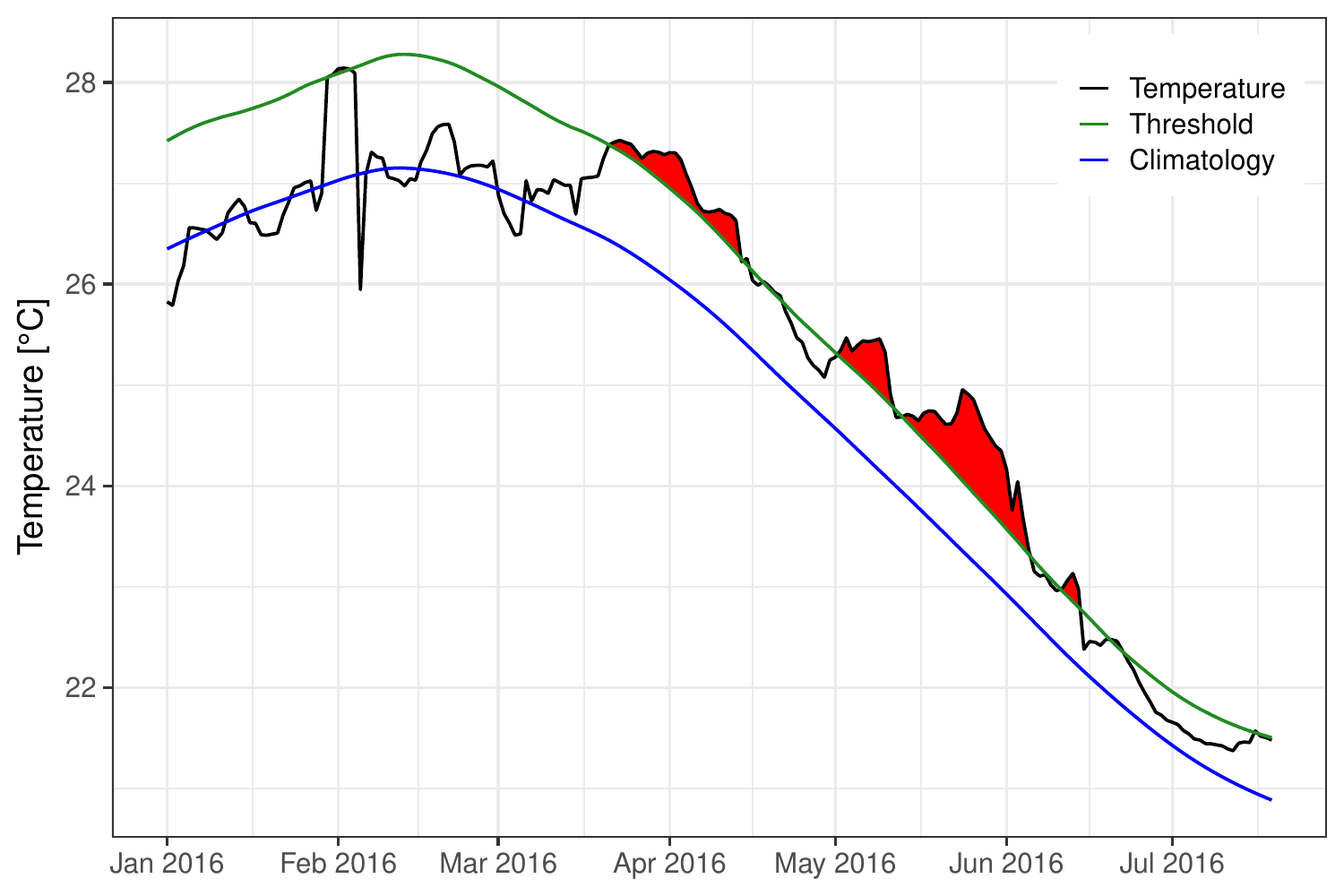}
    \caption{The SST climatology (blue), $90^{th}$ percentile MHW threshold (green), and SST time series (black) for each MHW at south of $22 ^\circ$ S to Hervey Bay. The red filled areas indicate the period of time associated with the identified MHWs.}
    \label{fig:mhw_episodes}
    \vspace{-1em}
\end{figure}

The corresponding MHW events allow us to form episodes $E_k$ and IMPIT indices of the form considered in Section \ref{sec_IMPIT}. For example, Figure \ref{fig:mhw_episodes} displays the SST climatology, $90^{th}$ percentile MHW threshold and SST time series for a typical MHW episode at a location off the Queensland coast (south of $22 ^\circ$ S in Hervey Bay). The red area between the black and green curves identifies the episode. The intensity function we have selected is the ``mean temperature anomaly during the MHW''. 

\subsubsection*{Fishery data \label{subsec_data_fish}}\vspace{0.5em}

\begin{figure}[ht]
    \centering
    \includegraphics[width=10cm]{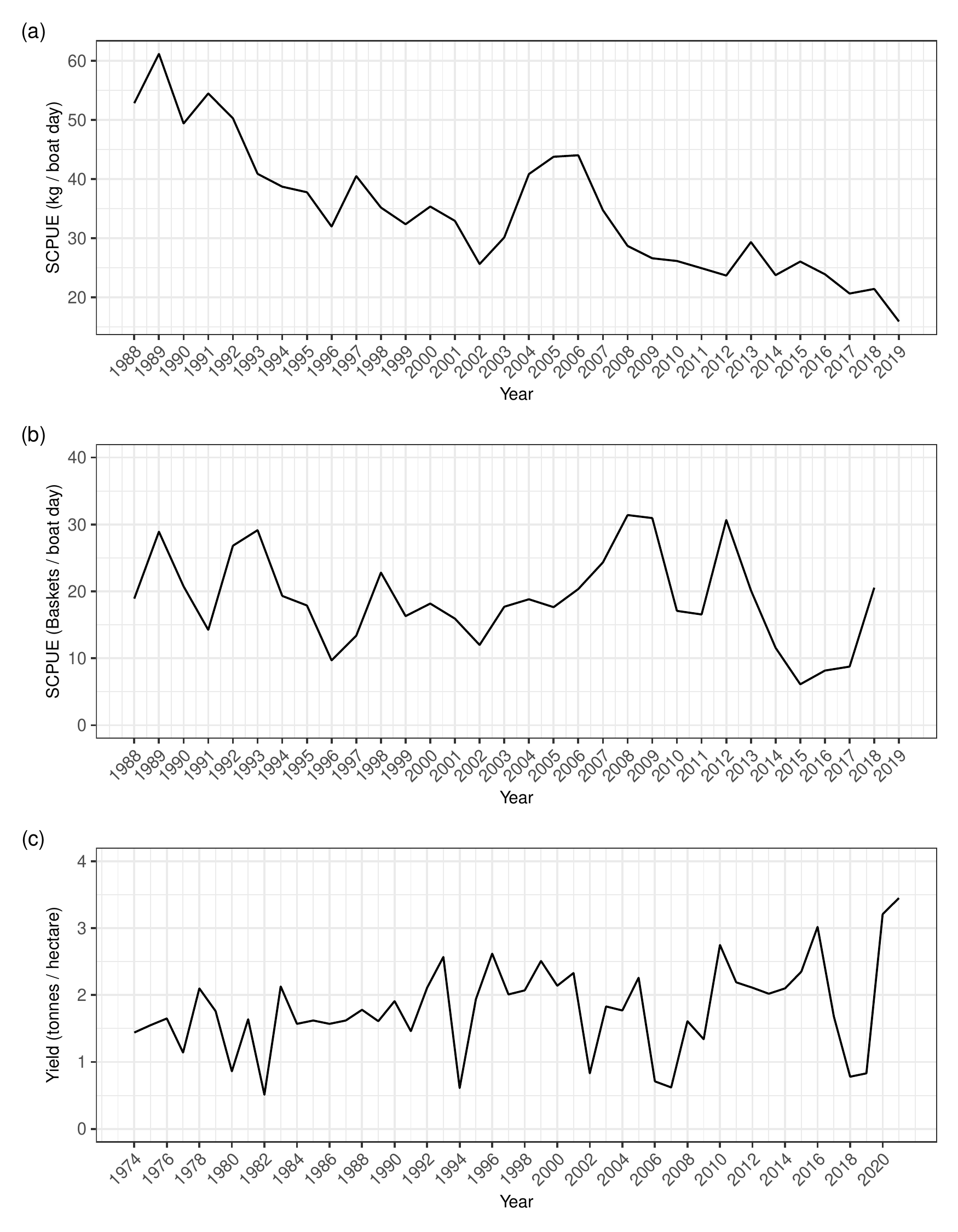}
    \caption{Time series of standardised catch rates for: (a) snapper ($1988-2019$) for the Queensland commercial line fishery, (b) saucer scallop in November ($1988-2018$) for the Queensland trawl fishery and (c) yield per hectare of wheat production in New South Wales, $1974-2021$.}
    \label{fig:scpue_series}
    \vspace{-1em}
\end{figure}

For snapper, standardised catch rate data were provided by Fishery Queensland. The top panel in Figure \ref{fig:scpue_series} displays the annual time series of SCPUE for snapper from the Queensland commercial line fishery. Overall, we observe a decreasing trend in catch rates across most of the dataset, following a peak in SCPUE in $1989$. In the case of scallop, catch rates were obtained from \citet{Wortmann_2020scallop}. Catch rates from November were chosen for use in the analysis because the fishing season has traditionally commenced in November, scallop catch rates generally peak at this time, and there is negligible fishing effort from May-October \citep{ONeill_2020stock}. The November catch rate has suffered a sharp decline in recent years. In Figure \ref{fig:scpue_series}b we observe a marked decline from $2012$ to $2017$ with a partial recovery in $2018$. Note that there are no November catch rate data after the fishery was closed in $2018$.

\clearpage

\subsubsection*{Agricultural data \label{subsec_data_wheat}}\vspace{0.5em}

Data on wheat production in NSW, from $1974$ to $2021$, were obtained from the Australian Government, Department of Agriculture, Fisheries and Forestry (DAFF) (\url{https://www.agriculture.gov.au/abares/research-topics/agricultural-outlook/data#australian-crop-report-data}). We have considered the wheat production in NSW as this is the main crop grown in the state, which is the second-highest producing State in Australia \citep{DPI_2007wheat}. In Figure \ref{fig:scpue_series}c we observe that even though there are many ups and downs, there is no strong trend in either direction. However, the low values in $1982$, $1994$, $2007$ and $2018-2019$ coincide with strong and moderate El Niño episodes (\url{http://www.bom.gov.au/climate/history/enso/}). There was an overall dry period across eastern and southeastern Australia over these two last years \citep{Wang_2020two}.

\subsection{The weak baseline associations}\vspace{0.5em}
There is a natural expectation that  environmental signals influence the abundance and hence harvest of both fish and crops \citep{Aburto_2010climatic, French_2021quantitative, Kangas_2022recovery, Potgieter_2002spatial, Wan_2022does, Zheng_2018value}. 
Indeed, marine heatwaves can have a big impact on coastal fisheries. For example, \citet{Caputi_2015management} report that adult biomass of saucer scallop and blue swimmer crabs in Western Australia (WA) declined as a result of the MHW of $2010/2011$. As a consequence, these commercial fisheries were closed from $2012$ to $2016$.

Furthermore, there could be a link between MHWs and SOI oscillations \citep{Meynecke_2012the, Loughran_2017understanding, Sen_2020drivers}. For instance, the MHW of $2010$/$2011$, which was reported to be the most extreme event (in intensity, extent and duration) ever recorded in WA, was a consequence of a strong Pacific La Ni\~na episode \citep{Caputi_2015management, Molony_2021what}. Further connections between La Ni\~na and El Ni\~no episodes and harvest of fish and crops were discussed in \citet{Meynecke_2012the, Rimmington_1993forecasting, Gutierrez_2017impacts}. In particular, 
the influence of environmental variables on the abundance of certain Queensland fishery species had been reported in \citet{Courtney_2015physical, ONeill_2020stock, Filar_2021modelling}. 

It is also widely accepted that El Ni\~no occurrences adversely influence crop yields \citep{Iizumi_2014impacts, Gutierrez_2017impacts}. Specifically, they have negative impact on Australian wheat production \citep{Rimmington_1993forecasting, Yuan_2015impacts, Zheng_2018value, Potgieter_2002spatial}.

\begin{figure}[ht]
    \centering
    \includegraphics[width=11.5cm]{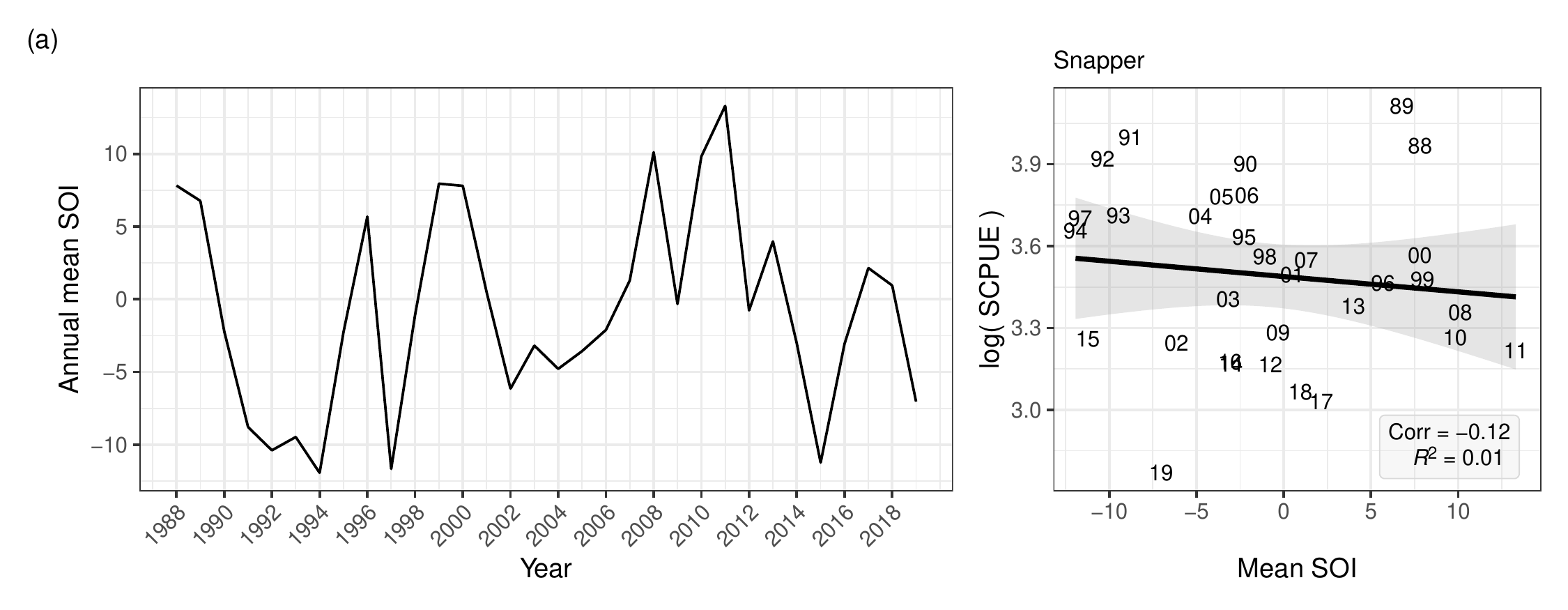}
    \includegraphics[width=11.5cm]{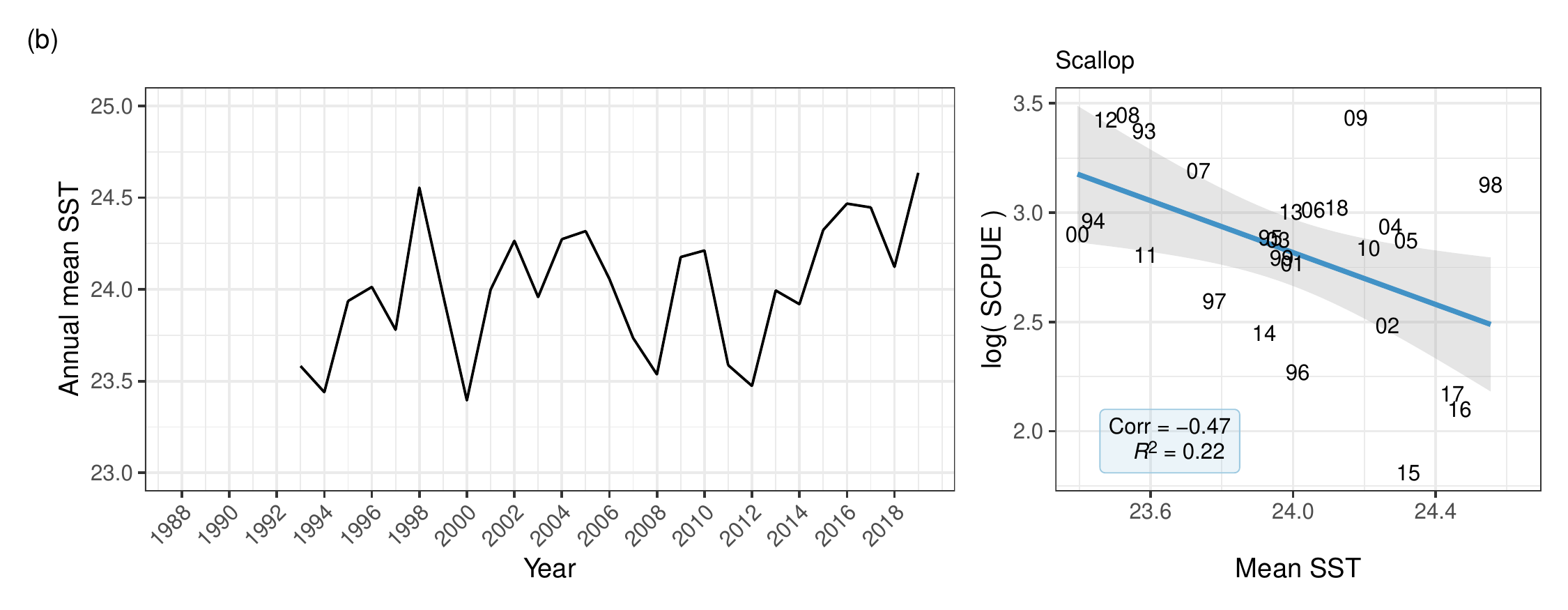}
    \includegraphics[width=11.5cm]{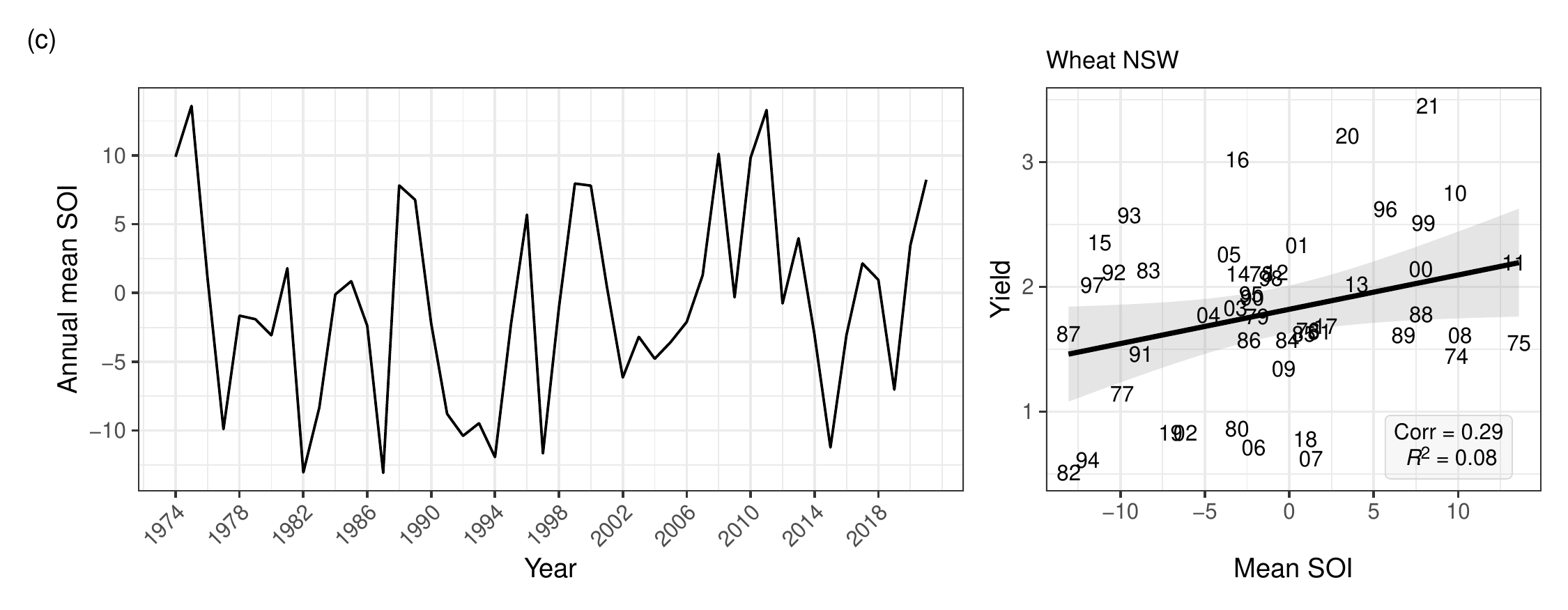}
    \caption{(a) Time series of annual mean SOI index $1988-2019$ (left) and scatterplot between annual mean SOI and annual log-transformed SCPUE of snapper (right). (b) Time series of SST annual mean $1993-2019$ in the saucer scallop fishery region (left) and scatterplot between SST annual mean and November log-transformed SCPUE of saucer scallop (right). (c) Time series of annual mean SOI index $1974-2021$ (left) and scatterplot between annual mean SOI and annual wheat yield in NSW (right). Regression line coloured according to p-value, in blue correlations with p-values $\leq 0.05$ and in grey with p-value $>0.05$.}
    \label{fig:annual_mean_signal_and_scpue}
    \vspace{-1em}
\end{figure}

However, the drivers of the Southern Oscillation phenomenon are complex, and as a result, there are no obvious trends in either the SOI signal, or its association with the catch rates of snapper or NSW wheat yield. The top and bottom right panels of Figure~\ref{fig:annual_mean_signal_and_scpue} show that there is no significant linear association between annual catch rates of snapper and wheat yield with the raw, unmodified annual mean SOI index. Still, there appears to be weak but significant negative association between annual mean sea surface temperature and the November catch rates of scallop as can be seen from the middle right panel of the same figure, that is consistent with earlier studies \citet{ONeill_2020stock}. The top and bottom left panels of Figure~\ref{fig:annual_mean_signal_and_scpue} indicate absence of linear trend in the mean annual SOI signal and the middle left panel indicates weak upward trend in the mean annual SST signal.

Nevertheless, it is possible that important associations are not revealed because the use of baseline annual means of the raw SOI and SST failed to account for the importance of persistence, recency, and the timing of the relevant intermittent episodes. Indeed, analyses of the remainder of this section illustrate, with the help of IMPIT indices, that this might be the case. These analyses should be viewed as exploratory rather than confirmatory.

\newpage

In the remainder of this section, we demonstrate that IMPIT indices can be calibrated to reveal previously hidden associations between SOI and SST signals and the catch rates of snapper and scallop in QLD and wheat yield in NSW. Relationships between the environmental variables and the fishery catch rates and wheat yield data were examined using correlation analyses. In the case of SOI, correlation analyses were based on $31$ years ($1988-2019$) when working with catch rate data and $47$ years ($1974-2021$) when considering yield data. In the case of SST, data were available only since $1993$, resulting in a shorter time series.
\newpage

\subsection{SOI threshold-crossing indices and snapper (\textit{Chrysophrys auratus}) \label{subsec_LNsnapper}}\vspace{0.5em}

We investigated linear associations between snapper annual SCPUE and a range of threshold crossing Super$X^{u} (8)$ IMPIT indices constructed using the methodology of Subsection \ref{subsec_threshold_cross}. The intensity of $E^u_k(8)$ episodes in Super$X^{u} (8)$ indices was computed by the Equation \eqref{Eq_I_av}.

\begin{figure}[ht]
    \centering
    \includegraphics[width=16cm]{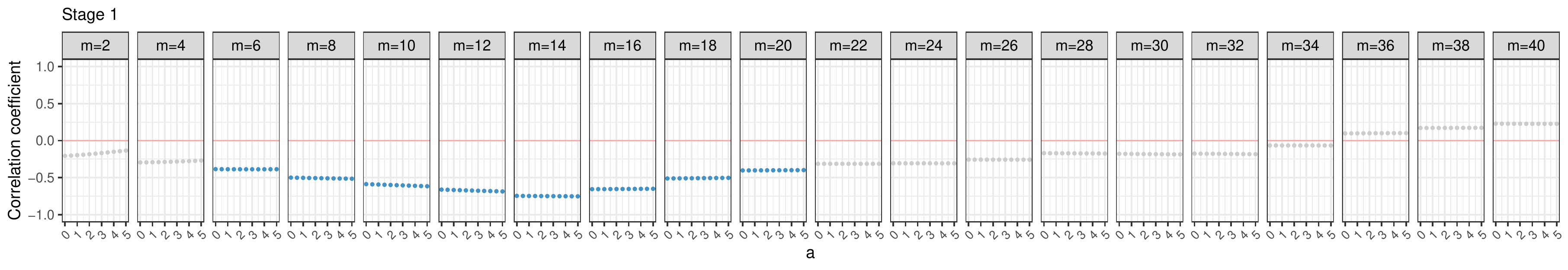}
    \includegraphics[width=16cm]{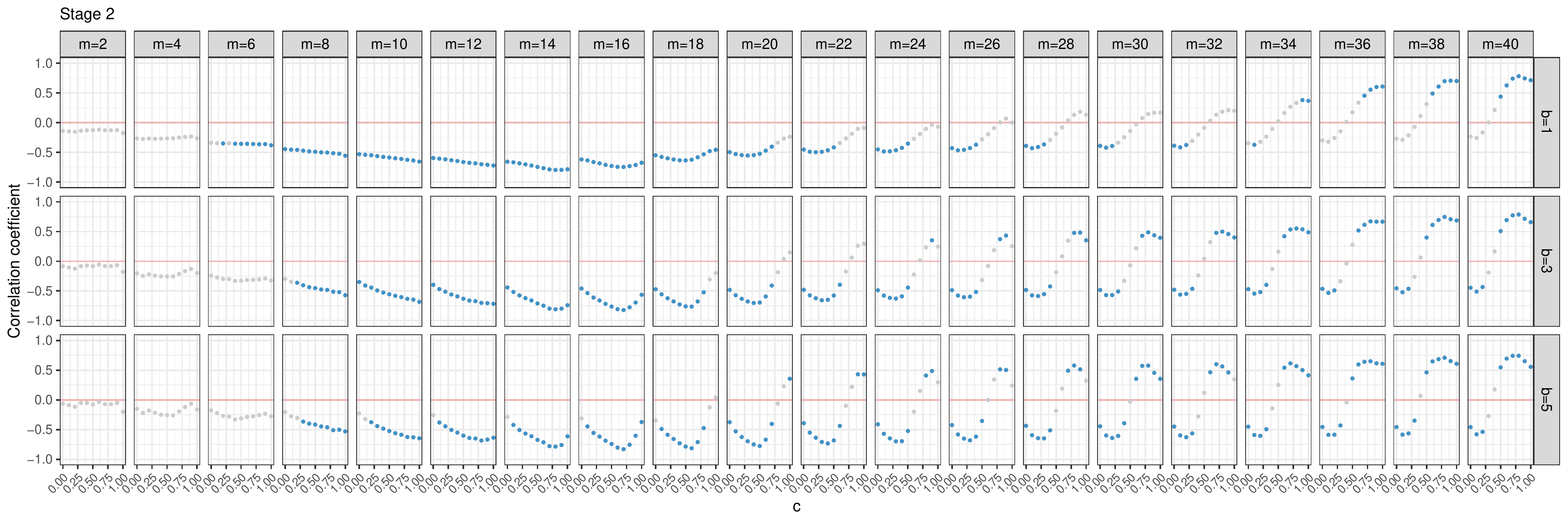}
    \includegraphics[width=16cm]{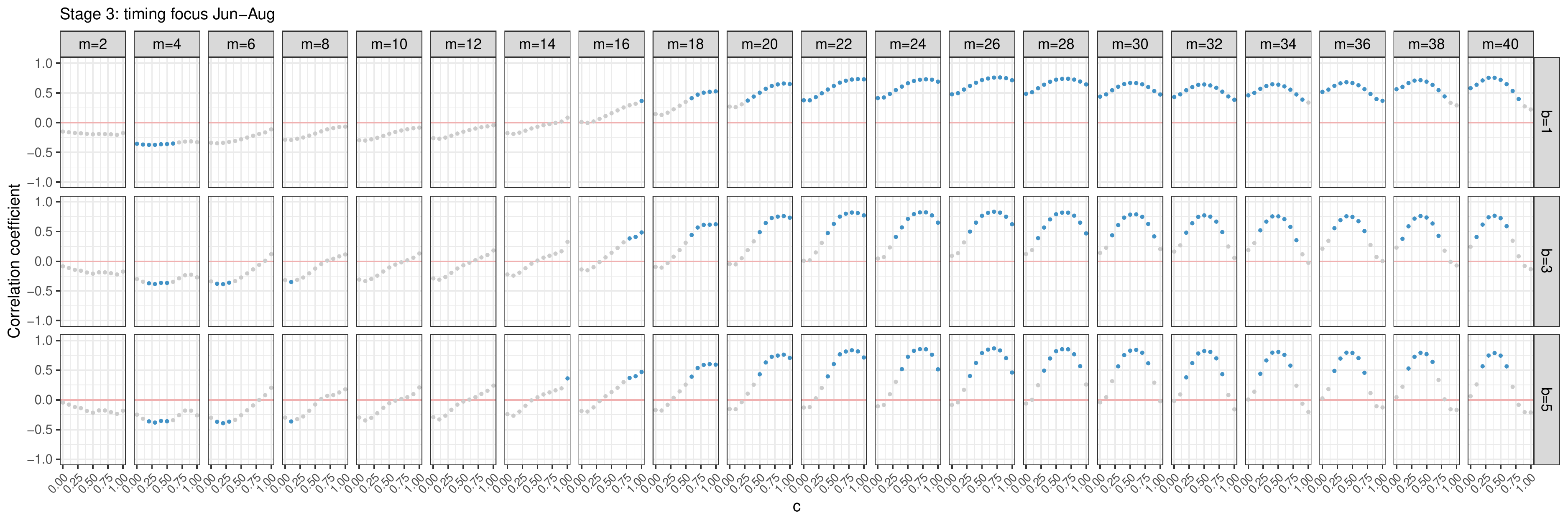}
    \caption{Correlation coefficient between Super$X^{u} (8)$ indices and annual snapper SCPUE ($1988-2019$). IMPIT indices were constructed for different values of \textit{memory} ($m$ in years), \textit{persistence} according to Equation \eqref{Eq_w1} with $a=2$, \textit{recency} conforming to Equation \eqref{Eq_w2} varying $b$ and $c$ and \textit{timing} according to Equation \eqref{Eq_w3} with $d=1$. Snapper peak spawning season (June-August) was the special season considered in Stage 3. Values coloured according to p-value, in blue scale correlations with p-values $\leq 0.05$ and in grey correlations with p-value $>0.05$.}
    \label{fig:snp_calib_stages}
    \vspace{-1em}
\end{figure}

We implemented the stage-wise explorative calibration described in Subsection \ref{subsec_calibration}. Figure \ref{fig:snp_calib_stages} depicts representatives of composite maps obtained at each stage, pruned for ease of display. For instance, we displayed only even values of the memory parameter $m$ which ranged from $1$ to $41$ years. The value of $41$ corresponded to the maximum longevity of snapper in the eastern coast stock reported in \citep{Wortmann_2018snapper}

The top panel of Figure \ref{fig:snp_calib_stages} corresponds to the first stage. Setting $w_2=w_3=1$, it displays plots of the correlation coefficient versus the parameter $a \in [0,5]$ for each value of $m$. Those plots exhibited a roughly linear pattern with very small variation. Hence we chose $a=2$, which corresponds to the middle pink curve for $w_1$ in Figure~\ref{fig:w1}.

\newpage 

At the second stage, while setting $w_3=1$ (and $a=2$) we plotted the correlation coefficient versus the parameter $c$ for alternative pairs of $m$ and $b$. The third stage considers the overlap of $E^u_k(8)$ episodes with the snapper peak spawning season of June-August and generates composite maps for alternative values of $d$. The second and third top panels of Figure \ref{fig:snp_calib_stages} display the resulting representative composite maps. The bottom panel shows the map for only $d=1$, which led to promising candidate configurations. For simplicity, only results for values $b \in \{1, 3, 5\}$ are displayed. 

Careful examination of Figure \ref{fig:snp_calib_stages} reveals interesting patterns generated by the stage-wise calibration. In particular, the impact of increasing values of the memory parameter is notable. Moreover, the inclusion of the special (June-August) timing generated the most noticeable differences between stages as it led to the disappearance of all significant negative correlations for values of $m \ge 10$ years.

These patterns may also hint at the dual nature of the impact of high SOI values on snapper catch rates and potentially abundance. Focusing on $E^u_k(8)$ episodes overlapping winter spawning season is more directly related to reproductive success and subsequent fish abundance. Thus significant positive correlations may be expected here. 

On the other hand, the presence of significant negative correlations, for shorter memories at Stage 2, suggests that the impact of high SOI values during non-spawning times of year, may be associated with catchability (e.g., windier, wetter-than-average conditions depressing catch rates).

We note that each single dot in the composite maps of Figure \ref{fig:snp_calib_stages} corresponds to a unique configuration of parameters $m,a,b,c$ and $d$. A simple rule-of-thumb rule for identifying a  promising candidate configuration is to: (a) search for a dot with a high absolute value of the correlation and (b) one that maintains a high correlation value to its immediate left and right within its local panel plot, across neighbouring local panels to its left and right (nearby $m$ values), across neighbouring local panels above and below (nearby $b$ values) and also checks for nearby $a$ and $d$ values\footnote{Such checks can be easily performed within our app IMPIT$-a$.}.

In all likelihood, there will be multiple promising parameter configurations. Below, we discuss just one such configuration with parameters: $m=26$, $a=2$, $b=3$, $c=0.75$ and $d=1$. Naturally, users can choose one or more configurations that best suit their case study and research objectives. 
\newpage

\begin{figure}[ht]
    \centering
    \includegraphics[width=13.5cm]{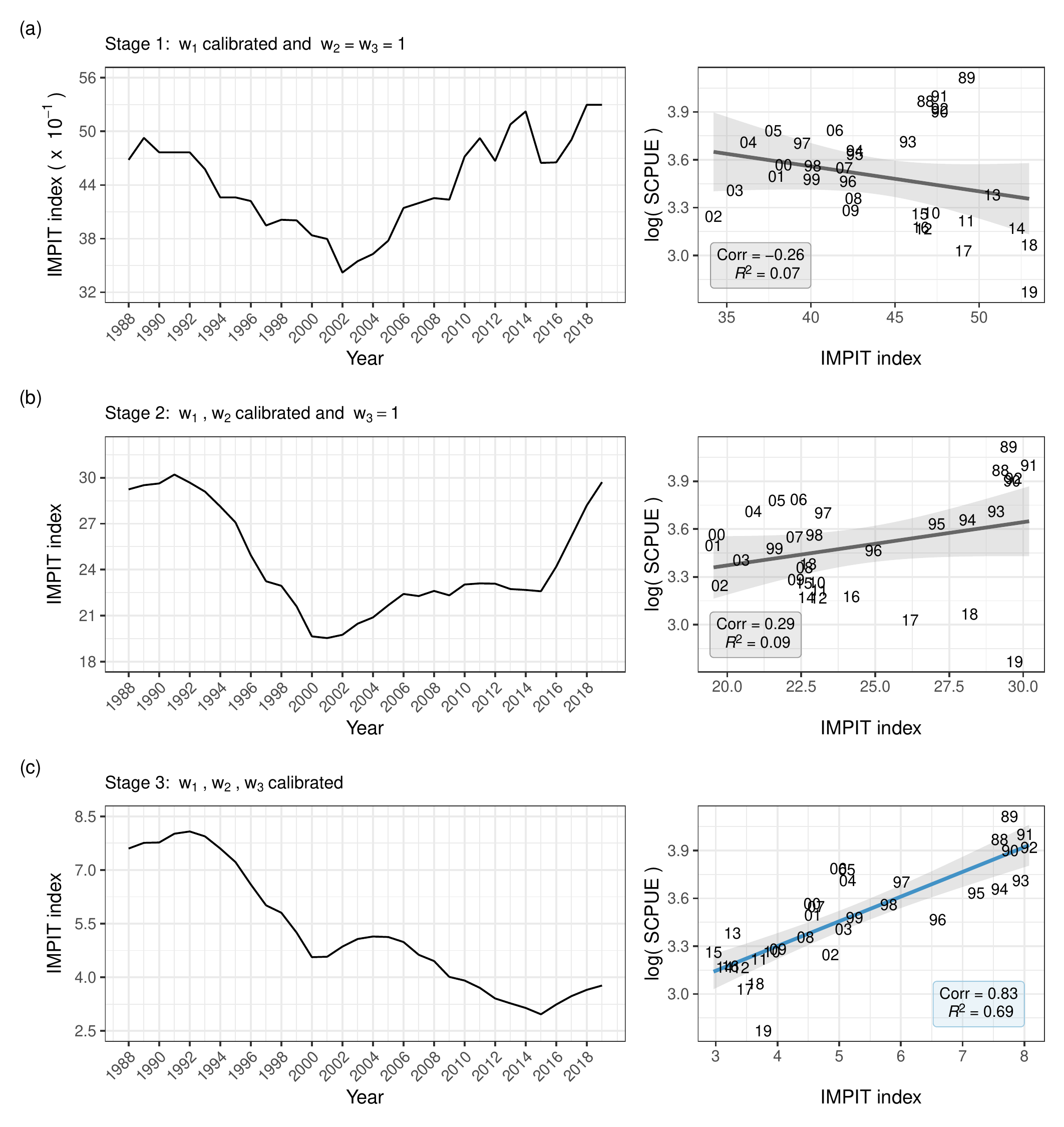}
    \caption{Time series of Super$X^{u}(8)$ indices and their corresponding scattergrams with annual snapper log-transformed catch rates (1988-2019). For IMPIT indices' construction, a memory of $26$ years ($m=26$) was used. The intensity of episodes was computed following the Equation \eqref{Eq_I_av}. Each row corresponds to a step of the \textit{stage-wise explorative calibration} of the relative importance weights. Leading to a parameter configuration: $a=2$, $b=3$, $c=0.75$ and $d=1$. Regression line coloured according to p-value, in blue correlations with p-values $\leq 0.05$ and, in grey otherwise.}
    \label{fig:snp_impit_scatters_m26}
    \vspace{-1em}
\end{figure}

The left panels of Figure \ref{fig:snp_impit_scatters_m26} exhibit the time series of Super$X^{u} (8)$ IMPIT indices correspond to the three calibration stages. From top to bottom, we observe marked changes in the shape of the resulting IMPIT index. At the top level (Stage 1) there appears to be essentially no linear trend over the $1988-2019$ period. Which continues to be the case at second level (Stage 2)\footnote{However, in both stages 1 and 2 there is an observable non linear U-shaped trend.}. As we move to the third level, we see that the Super$X^{u} (8)$ index exhibits a stronger downward linear trend. This reflects the importance of restricting to only the winter spawning season.

The right panels of Figure \ref{fig:snp_impit_scatters_m26} show the corresponding scattergrams of Super$X^{u} (8)$ indices with annual snapper log-transformed catch rates. These illustrate statistically significant changes in both the magnitude of the correlations and the degree of goodness of fit of straight line trends. 

In particular, note that at Stage 1 (top row), there is a non-significant negative linear association between the Super$X^{u} (8)$ index and the catch rate. Inclusion of the calibrated recency weight, at Stage 2 (second row), leads to a positive but still non-significant correlation of $0.29$. However, considering the winter spawning season at Stage 3 (third row), leads to a much higher correlation of $0.83$ corresponding to $R^2$ of $0.69$. 

Our illustrative analyses indicate that Super$X^{u} (8)$ index with a memory of $26$ years and suitably calibrated parameters exhibits some remarkably strong linear associations with log-transformed annual snapper catch rates. The value $c=0.75$ in the selected parameter configuration shows that episodes of approximately $20$ ($26 \times 0.75$) years in the past receive the highest recency importance weight ($w_2=1$). Episodes starting between $15$ and $20$ years in the past receive recency weights increasing from $0.79$ to $1.0$, while those starting between $20$ and $25$ years have recency weights decreasing from $1.0$ to $0.59$. The large increase in correlation from Stage $2$ to Stage $3$  suggests an impact of Super$X^{u} (8)$ on spawning and possibly also on subsequent abundance. In summary, the results suggest episodic events characterised by elevated ($ \ge 8$) SOI values during the June-August peak spawning season during the $15$-$25$ years prior to catch are associated with elevated snapper commercial catch rates.

\subsection{Marine Heatwave indices and saucer scallop (\textit{Ylistrum balloti}) in Queensland \label{subsec_MHWscallop}}\vspace{0.5em}

For saucer scallops, linear associations between standardised catch rates and a range of MHW IMPIT indices were explored using the methodology of Section \ref{sec_IMPIT}. As with snapper, the intensity of MHW episodes was computed by Equation \eqref{Eq_I_av}. We chose to examine the MHW index because of its association with other indices, its influence on scallops and a relatively long time series of available data.

Above average winter SSTs (June-August) are known to be negatively correlated with November-January scallop catch rates from 1988-2016, as reported in \cite{ONeill_2020stock}. As noted in \cite{Caputi_2015management}, in WA annual recruitment of saucer scallops is also correlated with SST (see also \citet{Joll_1995environmental, Lenanton_2009ongoing}). In addition, an extreme MHW in the summer of $2010$-$2011$ had catastrophic impact on the WA stock \citep{Caputi_2014catch, Caputi_2015management, Caputi_2019factors}. Scallops have an extended spawning season from April to October. However, between April and May most scallops are already sexually mature \citep{Dredge_1981reprod}. Hence, we shall refer to April-May egg production as the ``autumn component of the spawning season''. 

\begin{figure}[ht]
    \centering
    \includegraphics[width=12.5cm]{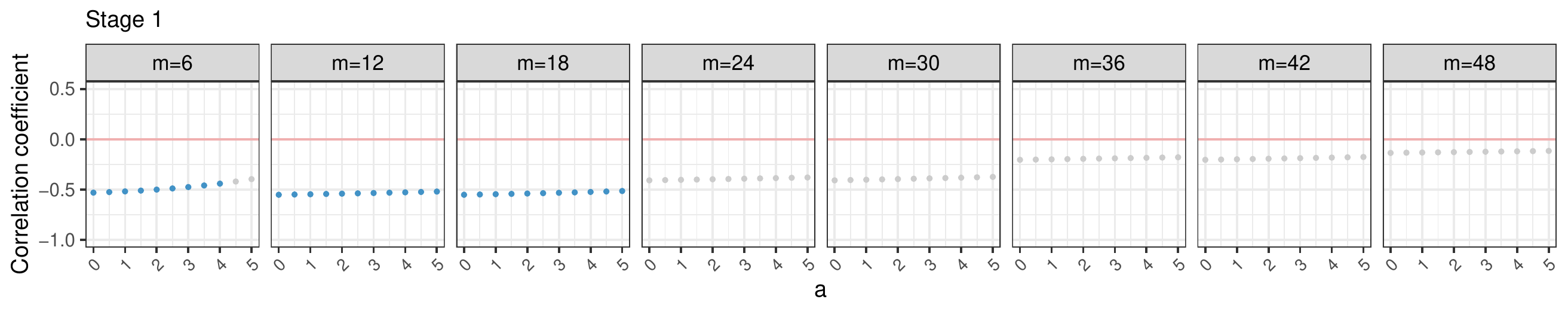}
    \includegraphics[width=12.5cm]{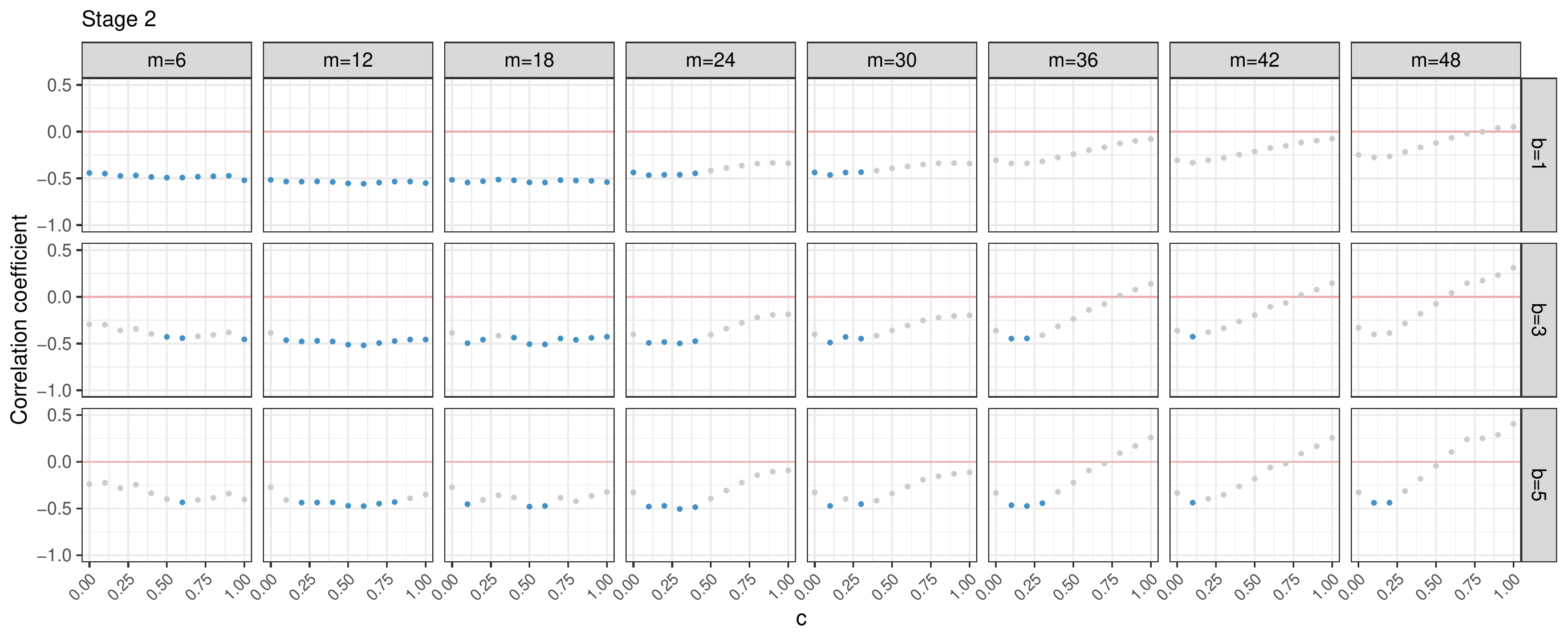}
    \includegraphics[width=12.5cm]{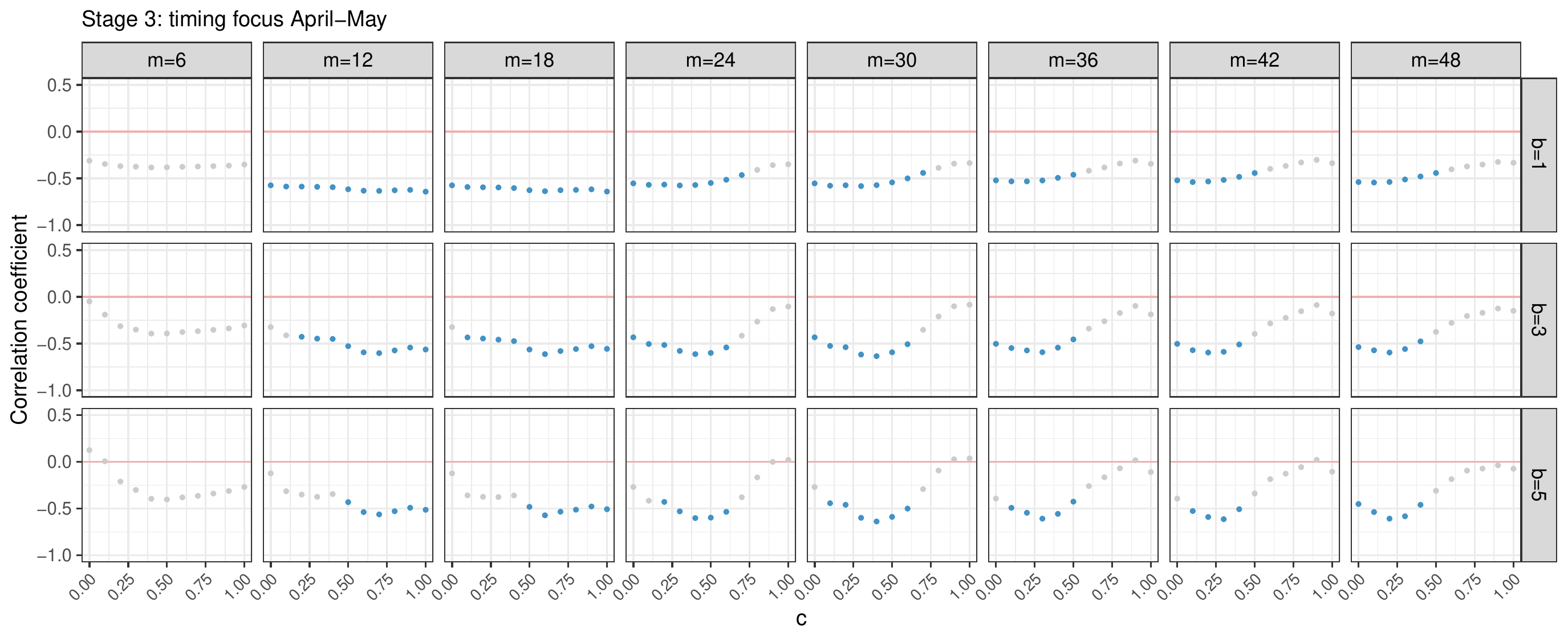}
    \caption{Correlation coefficient between MHW IMPIT and November scallop SCPUE ($1997-2018$). IMPIT indices were constructed for different values of memory ($m$ in months), \textit{persistence} according to Equation \eqref{Eq_w1} with $a=0$ and \textit{recency} conforming to \eqref{Eq_w2} varying $b$ and $c$ and \textit{timing} according to Equation \eqref{Eq_w3} with $d=1$. April-May was the special season considered in Stage 3. Values coloured according to p-value, in blue scale correlations with p-values $\leq 0.05$ and in grey correlations with p-value $>0.05$.}
    \label{fig:scp_calib_stages}
    \vspace{-1em}
\end{figure}

We performed the stage-wise explorative calibration described in Subsection \ref{subsec_calibration}. Figure \ref{fig:scp_calib_stages} depicts representatives of composite maps obtained at each stage. For the memory parameter $m$, we considered a $48-$month window. This was chosen because scallops can in some instances live for up to $4$ years \citep{Dredge_1985estimates, Courtney_2022estimating}. Figure \ref{fig:scp_calib_stages} shows values of memory increasing in steps of six months. 

The top panel of Figure \ref{fig:scp_calib_stages} corresponds to the first stage, Setting $w_2=w_3=1$, it displays plots of correlation coefficient versus the parameter $a \in [0,5]$ for each value of $m$. Those plots exhibited a roughly linear pattern but in this case there was more variation compared to the case of snapper. Therefore, we chose to use $a=0$ in $w_1$, which means that when constructing indices, all episodes were assigned the same persistence weight.

At the second stage, while setting $w_3=1$ ($a=0$) we plotted the correlation coefficient versus the parameter $c$ for alternative pairs of $m$ and $b$ parameters. At the third stage, we repeated the second stage with only those episodes that overlap the April-May component of the scallops spawning season. Second and third panels of Figure \ref{fig:scp_calib_stages} display the resulting representative composite maps. For simplicity, as in the case of snapper, only results for $b \in \{1,3,5\}$ are displayed. 
At the third stage, we generated composite maps for alternative values of $d$. Figure \ref{fig:scp_calib_stages} only displays the map for $d=1$ as it leads to promising candidate configurations.

Careful examination of plots in Figure \ref{fig:scp_calib_stages} reveals interesting patterns generated by the stage-wise calibration. In particular, the impact of considering the special timing (April-May) generated noticeable differences between stages for values of $m \geq 24$ months. 

\newpage
Typically, there will be multiple promising parameter configurations. Below, we discuss just one such configuration with parameters: $m=30$, $a=0$, $b=3$, $c=0.4$ and $d=1$. Naturally, users can choose one or more configurations that best suit their case study and research objective. 

\begin{figure}[ht]
    \centering
    \includegraphics[width=12.0cm]{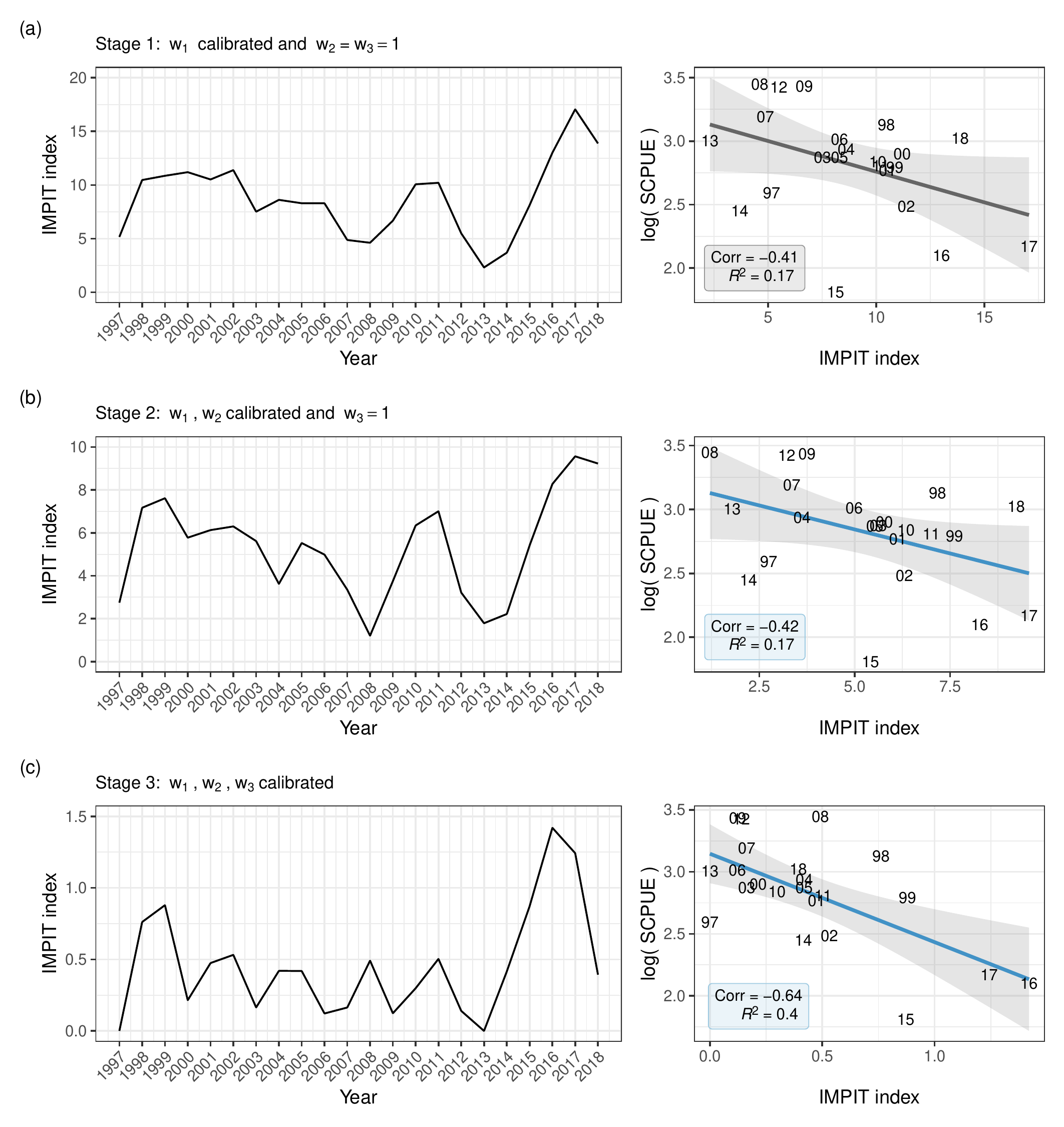}
    \caption{Time series of MHW IMPIT indices and their corresponding scattergrams with November scallop catch rates ($1997-2018$). For IMPIT indices' construction, a memory of $30$ months ($m=30$) was used. The intensity of episodes was computed following the Equation \eqref{Eq_I_av}. Each row correspond to a step of the \textit{stage-wise explorative calibration} of the relative importance weights. Leading to a parameter configuration: $a=0$, $b=3$, $c=0.4$ and $d=1$. Regression line coloured according to p-value, in blue correlations with p-values $\leq 0.05$ and, in grey otherwise.}
    \label{fig:scp_impit_scatters_m30}
    \vspace{-1em}
\end{figure}

The left panels of Figure \ref{fig:scp_impit_scatters_m30} exhibit the time series of MHW IMPIT indices for each calibration stage. From top to bottom, we observe that the shape of the resulting index changes. At top level (Stage 1) there appears to be a rough downward trend until $2013$ followed by an upward swing. That trend is less pronounced at second level (Stage 2). However, when we move from second to third level, we observe a period of relative stability between $2000$ and $2013$ followed by a marked increase. This reflects the effect of the addition of the spawning May-April timing into the calibration.

The right panels of Figure \ref{fig:scp_impit_scatters_m30} show the corresponding scattergrams of MHW IMPIT indices with November log-transformed scallop catch rates. In particular, note that at Stage 1 (first row), there is a negative linear association between MHW IMPIT index and the catch rate of $-0.41$. Inclusion of the recency weight calibration, at Stage 2 (second row), led to a very marginal but now statistically significant improvement of the correlation. Moreover, at Stage 3 (third row), the calibration of the timing $w_3$ weight, led to a highly significant correlation of $-0.64$ as compared to Stages 1 and 2, which accounts for $40 \%$ of variability.

Our illustrative analyses indicate that a MHW IMPIT index with a memory of $30$ months and suitably calibrated parameters exhibits some strong linear associations with log-transformed November scallop catch rates. The value of $a=0$ in the selected parameter configuration means that all episodes receive equal persistence weight and the value of $c=0.4$ shows that episodes of $12$ ($30 \times 0.4$) months in the past receive the peak recency importance weight ($w_2=1$). Episodes starting between $6$ and $12$ months in the past receive recency weights increasing from $0.74$ to $1.0$, while those starting starting between $12$ and $18$ months have recency weights decreasing from $1.0$ to $0.79$.

The increase in correlation in Stage 3 compared to Stages 1 and 2 is suggestive of an impact of April-May (autumn component of the spawning season) MHW IMPIT index on subsequent abundance. 
This means that MHW episodes ranging from $6$ to $18$ months in the past may strongly depress scallop catch rates in the following year. This is likely a result of either direct heat stress to the scallops themselves, or possible indirect effects of warm waters affecting primary production (i.e., the phytoplankton and zooplankton that scallops feed on) and therefore reducing food availability for scallops, predators and disease, all of which can affect scallop abundance \citep{Courtney_2015physical, Richardson_2020state}.

The strong correlation ($-0.64$) of the MHW IMPIT index with log-transformed SCPUE is not surprising since the bottom panel of Figure \ref{fig:annual_mean_signal_and_scpue} (and findings of \citet{ONeill_2020stock}) indicates a significant association with the raw SST signal. However, focusing on MHW IMPIT indices overlapping April-May reveals even stronger associations.

\subsection{SOI threshold-crossing indices and New South Wales wheat yield \label{subsec_NSWwheat}}\vspace{0.5em}

We investigated linear associations between annual yield per hectare of wheat and a range of threshold crossing Sub$X^{d} (-8)$ IMPIT indices constructed using the methodology of Subsection \ref{subsec_threshold_cross}. The intensity of $E^d_k(-8)$ episodes in Sub$X^{d} (-8)$ indices was computed by the Equation \eqref{Eq_I_av}.

\begin{figure}[ht]
    \centering
    \includegraphics[width=13.0cm]{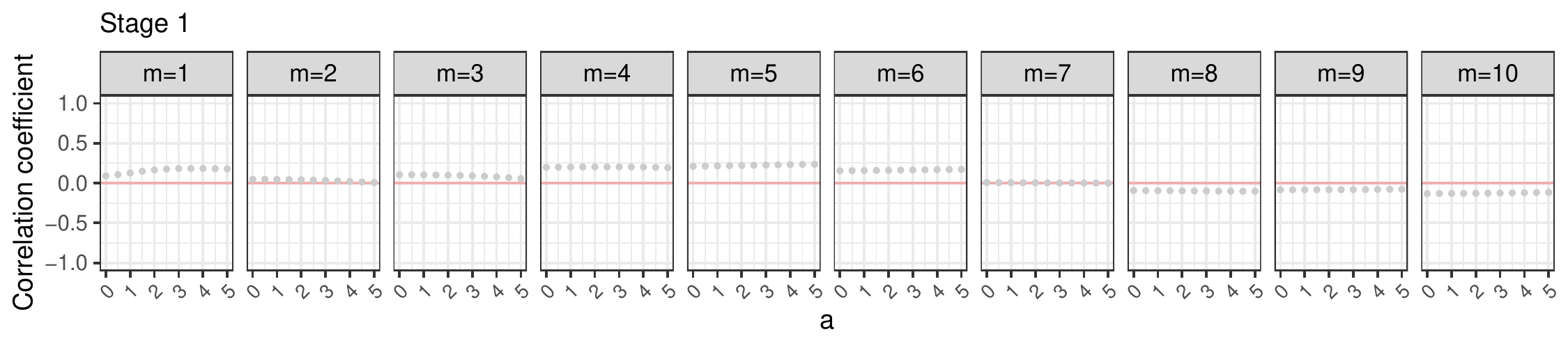}
    \includegraphics[width=13.0cm]{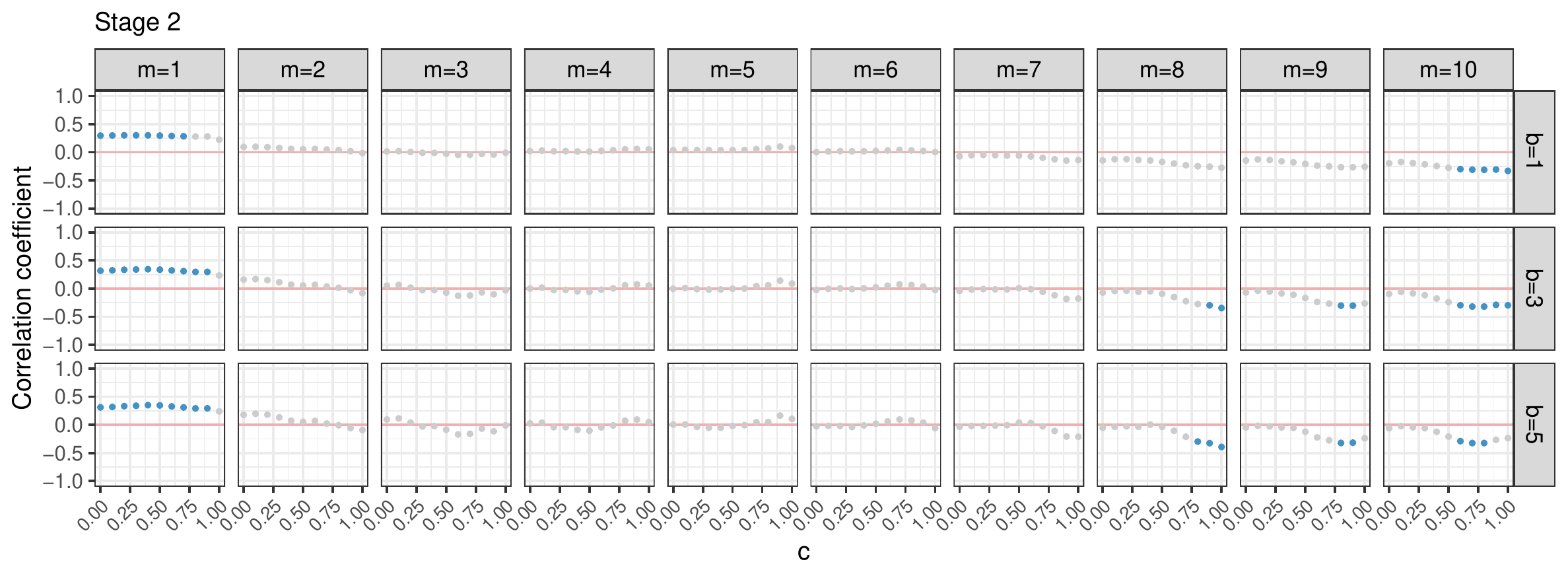}
    \includegraphics[width=13.0cm]{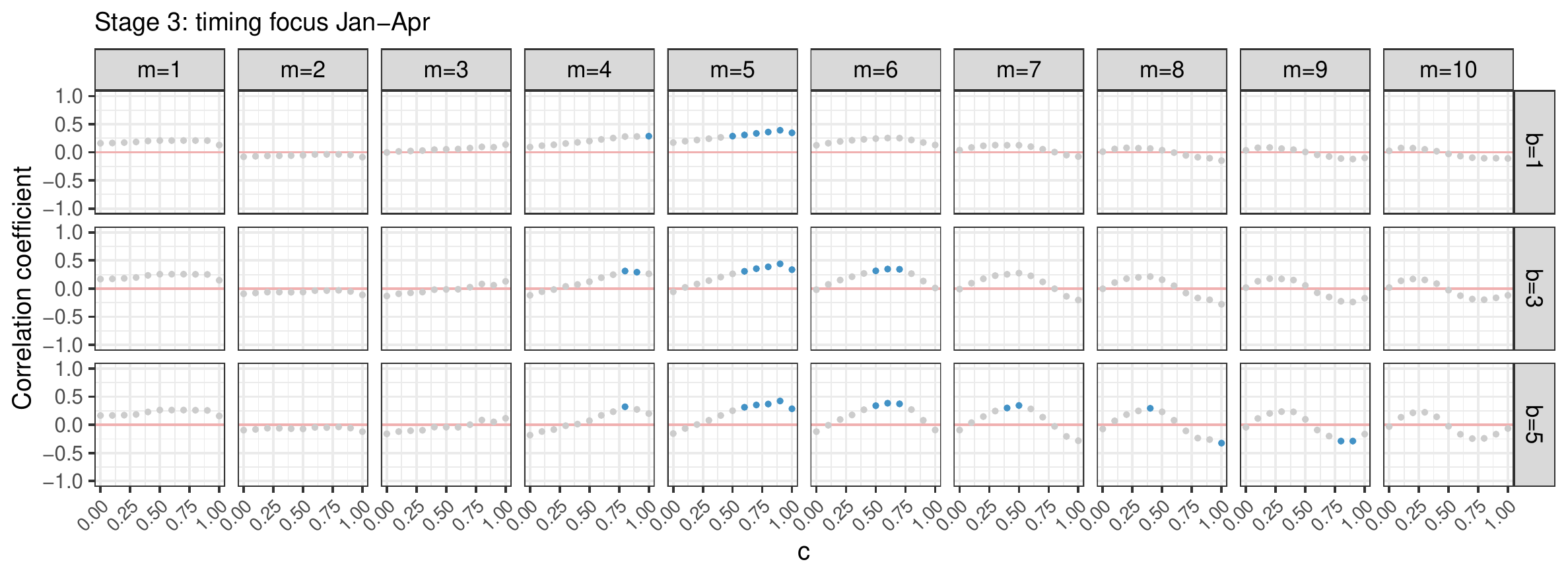}
    \caption{Correlation coefficient between Sub$X^{d} (-8)$ indices and NSW annual wheat yield ($1974-2021$). IMPIT indices were constructed for different values of \textit{memory} ($m$ in years), \textit{persistence} according to Equation \eqref{Eq_w1} with $a=2$, \textit{recency} conforming to Equation \eqref{Eq_w2} varying $b$ and $c$ and \textit{timing} according to Equation \eqref{Eq_w3} with $d=1$. January-April was the special season considered in Stage 3. Values coloured according to p-value, in blue scale correlations with p-values $\leq 0.05$ and in grey correlations with p-value $>0.05$.}
    \label{fig:wheat_calib_stages}
    \vspace{-1em}
\end{figure}

We implemented the stage-wise explorative calibration described in Subsection \ref{subsec_calibration}. Figure \ref{fig:wheat_calib_stages} depicts representatives of composite maps obtained at each stage, pruned for ease of display. For instance, we displayed only values of the memory parameter $m$ which ranged from $1$ to $10$ years. The value of $10$ was selected to check for long term effects. In recent decades, Australia has seen a shift towards higher temperatures and lower winter rainfall, which has had significant effects on many farmers \citep{Hughes_2022modelling}.

The top panel of Figure \ref{fig:wheat_calib_stages} corresponds to the first stage. Setting $w_2=w_3=1$, it displays plots of the correlation coefficient versus the parameter $a \in [0,5]$ for each value of $m$. Those plots exhibited a roughly linear pattern with very small variation and no significant correlations. As in Section \ref{subsec_LNsnapper}, we set $a=2$.

At the second stage, while setting $w_3=1$ (and $a=2$) we plotted the correlation coefficient versus the parameter $c$ for alternative pairs of $m$ and $b$ (see the second panel of Figure~\ref{fig:wheat_calib_stages}). For simplicity, only results for values $b \in \{1, 3, 5\}$ are displayed. We note that for the short memory of $1$ year we observed sustained statistically significant correlations for several values of $c$ and $b$ ranging from $3$ to $5$. There were also some isolated significant correlations for longer term memories of $5$ years and above. We emphasize $m=5$ because these associations carry over also to Stage $3$ discussed below. 

The third stage considers the overlap of $E^d_k(-8)$ episodes with the season of January-April which precedes the usual sowing period in NSW. Arguably, the conditions of the soil at the time of sowing have an impact on the subsequent yield. We then generated the composite maps for alternative values of the parameter $d$. The third panel of Figure \ref{fig:wheat_calib_stages} displays the representative composite maps for only one value $d=1$, as other values of $d$ result in similar outputs. 

Careful examination of Figure~\ref{fig:wheat_calib_stages} reveals interesting patterns generated by the stage-wise calibration. The significant positive correlations are to be expected as it is generally believed that decreasing negative values of SOI maybe associated with higher wheat yields \citep{Wan_2022does}. The sporadic significant negative correlations for long memories of $8$ years and above are noted but viewed as possibly spurious because they are less stable with respect to parameter changes. Moreover, the inclusion of the special (January-April) timing generated the most noticeable difference between stages as it led to the significant positive correlations for the memory of $m = 5$ years. At the same time, the associations at the memory of $m=1$ year became statistically insignificant when the episodes were restricted to the special period. 

The strongest observed correlation between the Sub$X^d (-8)$ IMPIT index of SOI and the wheat yield per hectare was $0.44$ which accounts for nearly $20\%$ of the variability. This is much higher than the $0.29$ reported in panel (c) of Figure~\ref{fig:annual_mean_signal_and_scpue} for the corresponding correlation with simple mean SOI index. Once again, this illustrates that IMPIT indices can potentially identify meaningful associations in this context of agricultural application.

\section{IMPIT app \label{sec_IMPITapp}}\vspace{0.5em}

In order to facilitate the use of IMPIT indices, a software tool IMPIT$-a$ was developed in the R environment \citep{R_2021} utilising ``Shiny'' \citep{Shiny_2021}, an R package that provides a web framework for building web applications using R (\url{https://cran.r-project.org}). It is important to note that R is open-source and widely used in environmental studies \citep{Aparicio_2019web, Diaz_2021aire}. 

The IMPIT$-a$ shiny app code is available on GitHub (\url{https://github.com/manumendiolar/IMPIT_shiny}) and needs to be run into the R  or RStudio \citep{RStudio_2021} environment, following instruction on the GitHub page. Alternatively, it can be accessed via the shinyapps.io platform (see Section~\ref{sec:SoftAvaib}). Note that because the app was built in R, it can be run on Windows, Linux, or macOS systems. With IMPIT$-a$ deployed as a web application, it can be used regardless of operating systems, hardware, or other installed software, since it can be run via a web browser. The aims of  IMPIT$-a$ are to:
\vspace{-0.5em}
\begin{itemize}
    \itemsep-0.25em
    \item Provide a user-friendly interface for constructing IMPIT indices
    \item Provide a smooth workflow ranging from importing and exploring raw data to defining episodes,
    \item Allow users to choose from a menu of intensity and relative weight functions
    \item Visualise imported data, defined episodes, and constructed IMPIT indices.
\end{itemize}

The app is intended to be self-contained in the sense that all the instructions and definitions are embedded in help messages within the software. The main tasks of the app can be summarised in four steps:
\begin{enumerate}
    \item Import the environmental data.
    \item Define discrete episodes.
    \item Construct IMPIT index.
    \item Explore IMPIT index.
\end{enumerate}

\begin{figure}[ht]
    \centering
    \includegraphics[width=13.0cm]{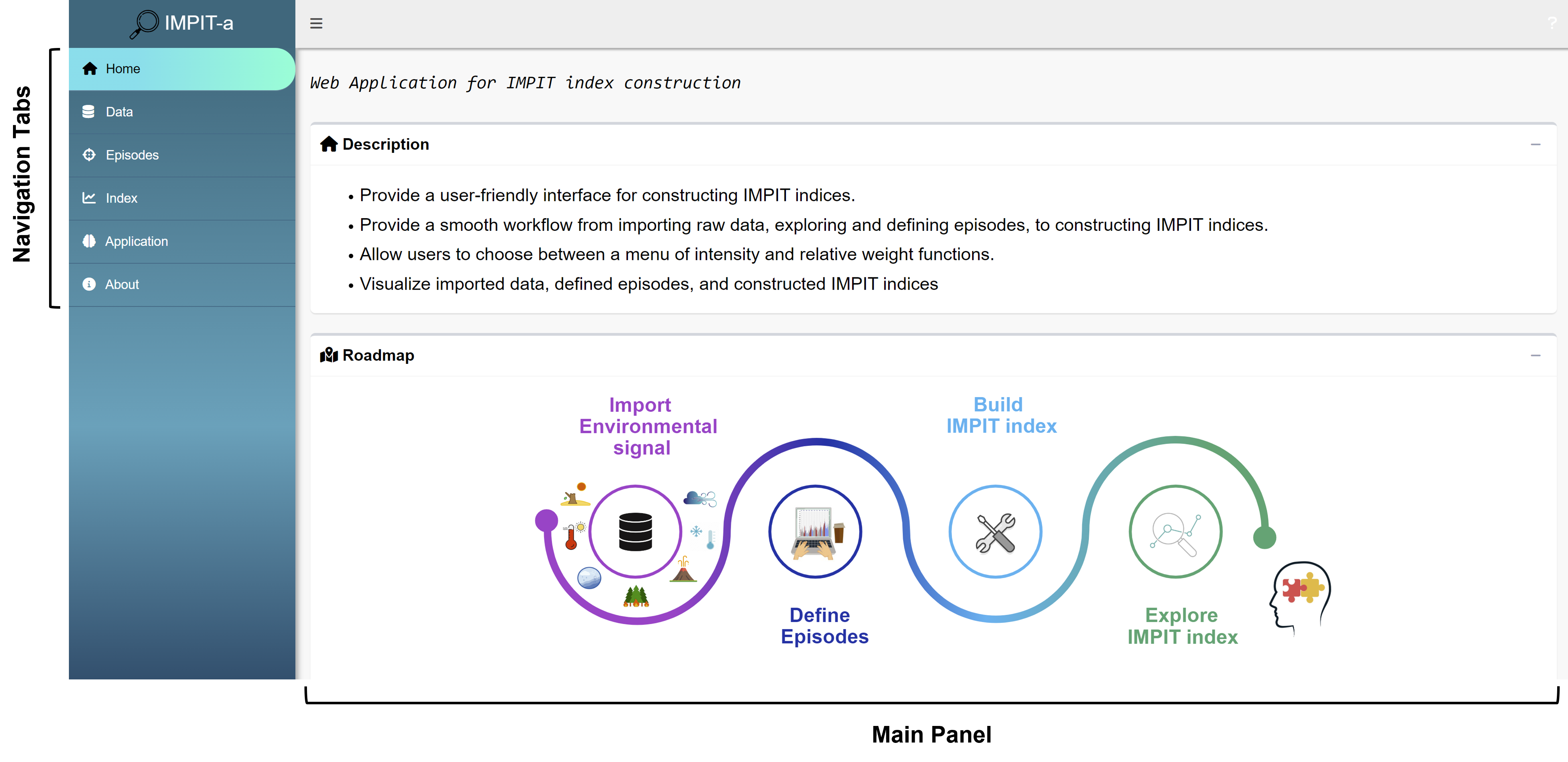}
    \caption{IMPIT$-a$ user interface structure. Sidebar Menu contains Navigational Tabs and app options. Main Panel contains app visualizations capabilities.}
    \label{fig:IMPIT_webapp_interface}
    \vspace{-2em}
\end{figure}

Each of the above steps is described in more detail in the remainder of this section. Figure \ref{fig:IMPIT_webapp_interface} shows the IMPIT$-a$ user interface structure with a dynamically linked sidebar menu and Main Panel. 

\newpage
\subsection{Input data\label{subsec_app_data_tab}}\vspace{0.5em}

Construction of IMPIT indices depends primarily on characterization of discrete episodes of interest. The app offers the options of either constructing threshold-crossing episodes or by directly uploading user-supplied episodes. 

\begin{figure}[!h]
    \centering
    \includegraphics[width=15cm]{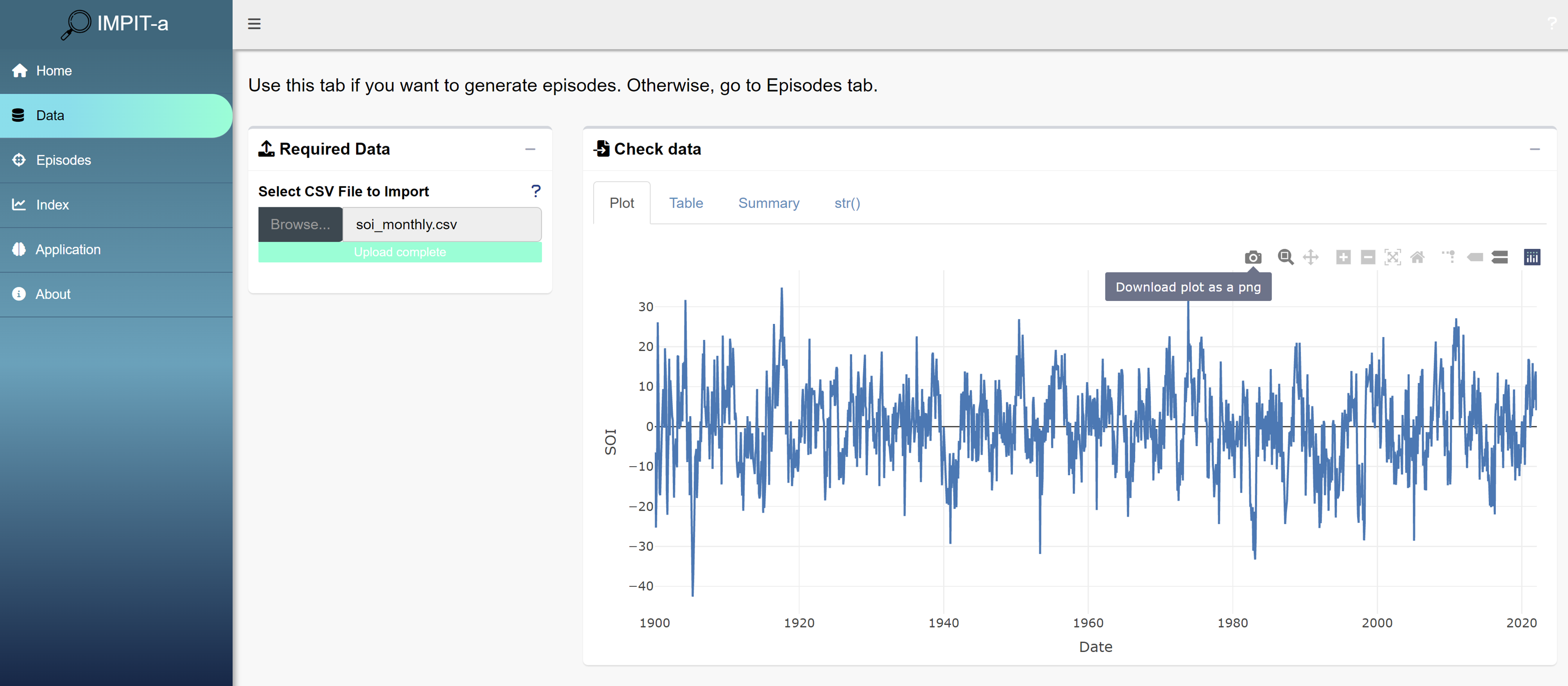}
    \caption{Example of Data tab, illustrated with time series plot of the monthly values of Southern Oscillation Index (SOI) from January 1900 to December 2020.}
    \label{fig:IMPIT_data_tab}
    \vspace{-1em}
\end{figure}

For the first option, the user has to upload the time series of the environmental signal, via the Data tab, and then construct the episodes, via the Episodes tab. Figure~\ref{fig:IMPIT_data_tab} shows an example of the Data tab. Uploaded data can be checked via Plot, Table, Summary and str() tab options. Note that all users can interact with graphs on display. Zooming in and out, point value display, panning graphs or saving the plot by clicking ``download plot as  a png'' button in the toolbar at the top of the graph. 

\begin{figure}[ht]
    \centering
    \includegraphics[width=15cm]{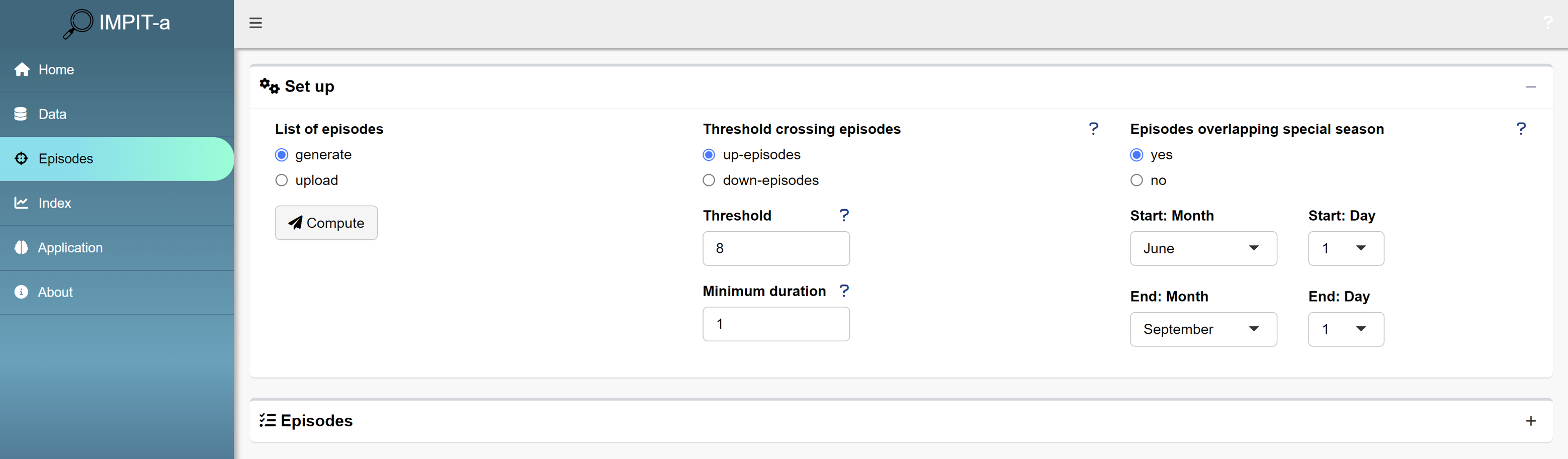}
    \includegraphics[width=15cm]{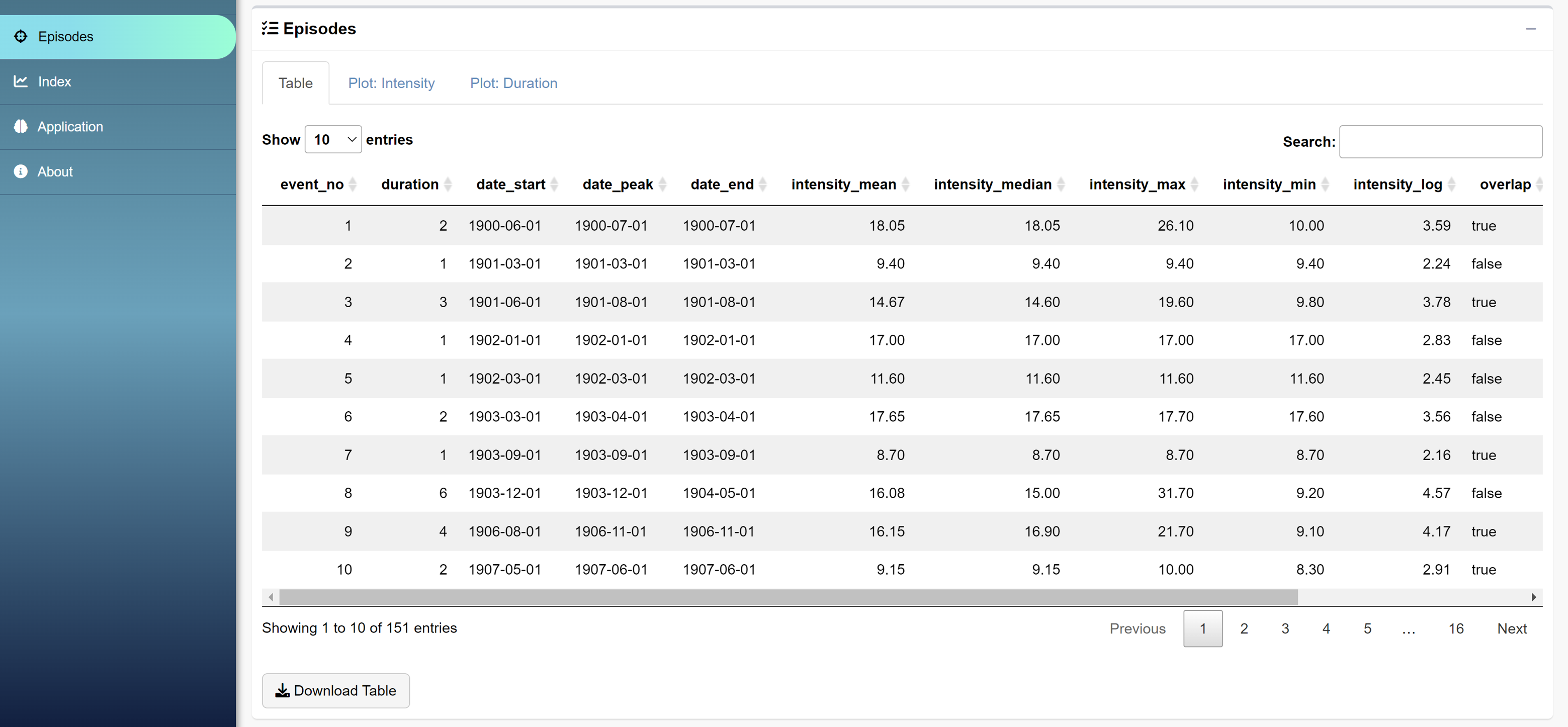}
    \includegraphics[width=15cm]{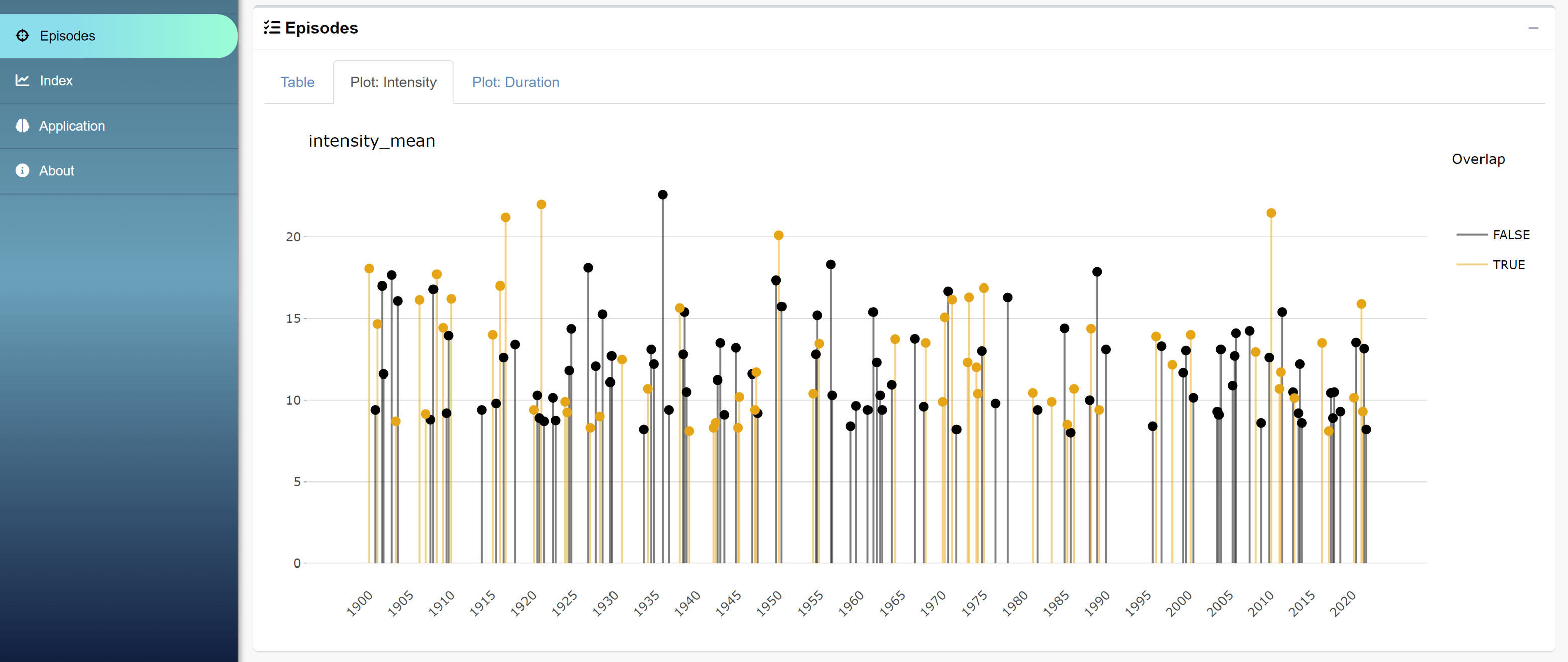}
    \caption{Example of Episodes tab capabilities. Top: In this case, $E^u_k(8)$ episodes were generated with timing focus on June-August from SOI signal (from January 1900 to December 2020). Middle: Table with detailed information of the episodes. Bottom: Lollipop chart showing the intensity mean of all episodes computed. In yellow the episodes that overlap the special season (June/01 - September/01).}
    \label{fig:IMPIT_epi_tab_generation}
    \vspace{-1em}
\end{figure}

Once this is done, the user can proceed to define episodes using the Episodes tab. This provides a definition of episodes based on the threshold-crossing indices introduced in Section \ref{subsec_threshold_cross}. User will have to select a threshold and choose between up or down episodes with respect to the chosen threshold. In addition, the minimum duration of episodes can also be selected (by default it is $1$). Minimum duration means, minimum consecutive values above or below the threshold (see Figure \ref{fig:IMPIT_epi_tab_generation}). However, if the user already has a list of episodes, they can be uploaded directly (see top panel of Figure \ref{fig:IMPIT_epi_tab_generation}). The app offers the option to  download the table with detailed information of episodes as well as lollipop charts showing the intensity mean and duration of all episodes. If the special season option was considered, episodes highlighted in yellow are those overlapping the special season (see bottom panel of Figure \ref{fig:IMPIT_epi_tab_generation}). 

\clearpage

\subsection{Index construction \label{subsec_app_index_tab}}\vspace{0.5em}

Once the episode list is generated or uploaded via the Episodes tab we are ready to build an IMPIT index. Suppose we consider up-episodes $E^{u}_{k}(8)$ and we want to build their corresponding Super$X^u(8)$ index. A crucial step is to set up a configuration of parameters associated with the memory, persistence, recency and timing of the episodes. Table \ref{tab:table_param_description} shows a summary of the main parameters, with their description, that can be found in the Index tab.

\begin{table*}[ht]
    \centering
    \caption{Summary description of IMPIT$-a$'s main parameters.}
    \label{tab:table_param_description}
    \begin{tabular}{p{1.5 cm} p{9cm}}
        \toprule
        \textbf{Parameter} & \textbf{Description}\\
        \midrule
        $m$ & \multicolumn{1}{m{9cm}}{Memory. A period of fixed and uninterrupted duration in the past with respect to the current observation in time.}\\
        \hline
        $a$ & \multicolumn{1}{m{9cm}}{Associated with the persistence of the episode. Dampens the rate of decay (Figure~\ref{fig:w1}). A value close to zero means that each episode will have nearly the same importance weight regardless of its duration.}\\
        \hline
        $b$ & \multicolumn{1}{m{9cm}}{Associated with the recency of the episode. Captures the rate of decay (Figure~\ref{fig:w2}) of the recency weight as the starting date deviates more from its peak. Low values flatten that relative importance weight.}\\
        \hline
        $c$ & \multicolumn{1}{m{9cm}}{Associated with the recency of the episode. Characterises the skewness of the recency weight and its peak (Figure~\ref{fig:w2}). A value close to zero indicates that recent episodes have higher weight than those starting late in the memory.}\\
        \hline
        $d$ & \multicolumn{1}{m{9cm}}{Associated with the timing importance weight of the episode. The weight is monotone increasing with $d$ (Figure~\ref{fig:w3}).}\\
        \bottomrule
    \end{tabular}
\end{table*}

\begin{figure}[ht]
    \centering
    \includegraphics[width=15.0cm]{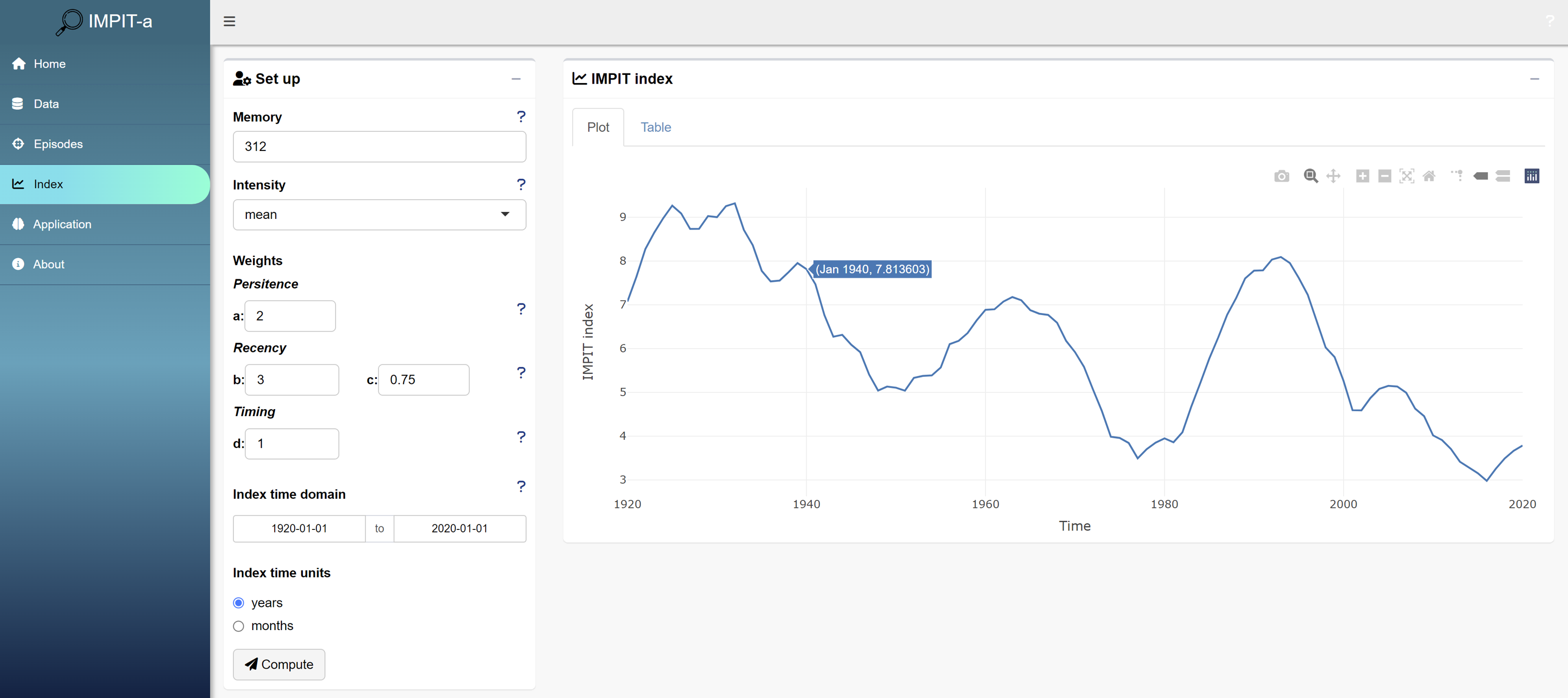}
    \caption{Example of Index tab to compute IMPIT index according to parameter's specifications. User can choose among a menu of memory, intensity and relative weight functions. Timing was considered and $E^u(8)$ episodes were weighted by their overlap duration with the June-August season. Note that memory units coincides with episodes units, in this case $312$ months ($26$ years).}
    \label{fig:IMPIT_index_tab}
    \vspace{-1em}
\end{figure}

In the next illustration, we consider a memory of $m=26$ years. This means that the value of the index in, for example, $2022$ will be computed based on the contribution of all $E^u(8)$ episodes from $1996$ until $2022$. For the intensity function, from the menu on this tab we can pick among the mean, median, minimum, maximum or logarithm of the episode's values. Suppose we choose the intensity function given by Equation \eqref{Eq_I_av}. 
Next, we have to decide on the relative importance weights of the episodes. Three aspects could be considered: persistence, recency and timing of episodes. For the persistence, let us consider $a=2$ which corresponds to the middle pink curve in Figure~\ref{fig:w1}. For the recency, suppose we choose $b=3$ for the dampening rate of decay and a value of $c=0.75$ which will be similar to the light green curve in the right panel of Figure~\ref{fig:w2}. This means that episodes starting recently will be more important than those starting late in the prescribed memory. Finally, for the timing, suppose episodes are weighted by their overlap duration with the special timing. To illustrate, we considered $d=1$ to calibrate the timing weight, which corresponds to dark orange curve (fifth from bottom to top) in the right panel of Figure~\ref{fig:w3}. Figure \ref{fig:IMPIT_index_tab} displays the resulting IMPIT index plot. User can download the table of IMPIT index values in the Table tab.  

\subsection{Output data exploration \label{subsec_app_app_tab}}\vspace{0.5em}

The ``Application'' tab was designed to provide a suite of tools for index exploration, including plot of IMPIT index with smoothing and the option to run a simple regression analysis between the index and a response variable of interest. The app's visualization functionality allows users to explore the index before exporting it for later use.

On the left panel (in the Application tab), users can upload an IMPIT index (generated with the app) and a response variable of interest. 
On the right panel, the first two tabs correspond to IMPIT index exploration, where user can visualise IMPIT index. A common approach to identify the structure of present trends, is to plot the time series in combination with a smooth fitted curve \citep{vonBromssen_2021toolbox}. 

The estimation of the smooth trend is usually achieved using a generalized additive model (GAM) or by LOESS smooths (\citet{Hastie_1986generalized, Wood_2017generalized, Cleveland_1979robust}. The app provides the option of explore IMPIT index with a LOESS smooth and its $95\%$ confidence intervals. 
The third and fourth tabs allow the users to check the uploaded response variable data. The fifth tab corresponds to the scatterplot between the IMPIT index and the response variable. And the sixth tab presents a summary of the regression analysis.

\section{Discussion \label{sec_Disc}}\vspace{0.5em}

In this paper, we developed a generic, parametrised family of weighted indices extracted from observations of a relevant environmental signal containing potentially important discrete episodes.
The methodology focuses on determining an appropriate memory length and assigning importance weights to  episodes that capture: intensity, persistence, intermittence and timing. 

We illustrated the effectiveness and possible uses of IMPIT indices in the context of fishery and agricultural applications. In particular, we considered standardised commercial fishery catch rate data from two fished species in Queensland, snapper (\textit{Chrysophrys auratus}) and saucer scallop (\textit{Ylistrum balloti}), and yield per hectare of wheat production in New South Wales. We then searched for associations between the trends in these data and IMPIT indices constructed from intermittent episodes embedded in SOI and SST signals.

Using SCPUE and yield per hectare of wheat as response variables, we showed that significant correlations exist with suitably calibrated IMPIT indices. Moreover, this occurred even though analyses using simple (baseline) means of the underlying environmental signals indicated no significant correlation in the case of snapper and wheat yield and only relatively weak correlation in the case of scallops. Hence, we demonstrated that our stage-wise parameter calibration of IMPIT indices plays a potentially important role in detecting associations between special seasons of either the studied species or wheat, and their environment. 

We also developed an IMPIT$-a$ software that expedites the index construction process and combines it with data which allows users to refine datasets/episodes and time periods analysed based on exploration, before exporting the resulting IMPIT index for further use. No specialized coding or expertise are needed to use IMPIT$-a$. The web interface ensures that the tool can be accessed from a web browser without installing any additional software. Although IMPIT$-a$ offers only a limited number of options for the intensity or weights functions, it can be extended to include other functional forms.

Identification of strong associations between a calibrated IMPIT index and response variables should be seen as a starting point for deeper follow-up investigations. These would seek to confirm that the corresponding IMPIT index is capturing a causative, not just correlative, relationship between the environmental signal and the study species.

IMPIT indices are quite generic and easy to interpret, and can be used across a wide range of environmental and ecological disciplines. The indices and their calibration should be seen as another tool in the toolbox of exploratory data analyses.

\section{Software availability}\label{sec:SoftAvaib}\vspace{0.5em}

\begin{itemize}
    \itemsep-0.25em
    \item [] Name of the software: IMPIT$-a$.
    \item [] Developer: Manuela Mendiolar.
    \item [] Contact Email: \texttt{m.mendiolar@uq.edu.au}.
    \item [] Tested browsers: Firefox and Google chrome.
    \item [] Software Required: R, RStudio.
    \item [] R-Packages required: shiny, shinydashboard, shinydashboardPlus, shinyFiles, shinyhelper, shinyalert, shinyvalidate, shinyjs, shinyWidgets, dashboardthemes, tidyverse, DT, plotly, spsComps and lubridate.
    \item [] Programming language: R version 4.1.2.
    \item [] Available since: 2022.
    \item [] The app source code is stored in a freely accessible GitHub repository hosted by Manuela Mendiolar: \url{https://github.com/manumendiolar/IMPIT_shiny}. 
    \item [] The app can be deployed via: \url{https://manumendiolar.shinyapps.io/impit_shiny/}.
\end{itemize}

\section{Declaration of competing interest}\vspace{0.5em}
The authors declare that they have no known competing financial interests or personal relationships that could have appeared to influence the work reported in this paper.

\section{Acknowledgments}\vspace{0.5em}
We would like to acknowledge Sue Helmke and Joanne Wortmann from the Queensland Government Department of Agriculture and Fisheries (DAF) for providing access to data. In addition, we are indebted to the entire team of the FRDC Project No. $2019-013$ for many valuable insights to the fishery applications.

\bibliographystyle{cas-model2-names}
\bibliography{cas-refs}

\end{document}